\documentclass[a4,12pt]{article}
\usepackage{amsmath,amssymb}
\usepackage{bm}
\usepackage{def_simbol}
\usepackage[dvipdfmx]{graphicx}
\usepackage{color}
\usepackage{cases}
\usepackage{comment}

  \makeatletter
    \renewcommand{\theequation}{%
    \thesection.\arabic{equation}}
    \@addtoreset{equation}{section}
  \makeatother

\usepackage[top=20truemm,bottom=20truemm,left=18truemm,right=18truemm]{geometry}

\definecolor{refkey}{rgb}{0.9451,0.2706,0.4941}
\definecolor{labelkey}{rgb}{0.9451,0.2706,0.4941}
  
\newcommand{\dsection}[1]{}
\renewcommand{\include}[1]{}

\usepackage[setpagesize=false,colorlinks=false]{hyperref}

\begin{document}

\begin{titlepage}
\begin{flushright}
OU-HET 867
\end{flushright}
\begin{center}

\vspace*{100truept}
{\huge The Enhancement of Supersymmetry in M-strings}\\ 
\vspace{30truept}
{ Yuji SUGIMOTO\footnote{sugimoto@het.phys.sci.osaka-u.ac.jp}\\ \vspace{2cm}{\em{Department of Physics, Graduate School of Science,\\
Osaka University, Toyonaka, 560-0043, Japan}}}\\ 
\begin{abstract}
We study two M5-branes on $A_{1}$ ALE space. We introduce some M2-branes suspended between the M5-branes. Then, the boundaries of M2-branes look like strings. We call them ``M-strings''. The M-strings have $\mathcal{N}=(4,0)$ supersymmetry by considering the brane configuration on $A_{1}$ ALE space. We calculate the partition function of M-strings by using the refined topological vertex formalism. We find that the supersymmetry of M-strings gets enhanced to $\mathcal{N}=(4,4)$ by tuning some K\"ahler parameters. Furthermore, we discuss another possibility of the enhancement of supersymmetry which is different from the above one. 
\end{abstract}
\end{center}
\end{titlepage}

\clearpage
\hrulefill
\tableofcontents
\hrulefill

\section{Introduction}
M5-branes are mysterious objects in M theory. The low energy theories on coincident multiple M5-branes are 6 dimensional $\mathcal{N} = (2,0)$ theories whose Lagrangians are not known. Then, one of the methods whom we can use to analyze is the duality.
\par

We consider two M5-branes on $A_{1}$ ALE space. When two M5-branes are separated by introducing some M2-branes, we can see some 2 dimensional objects in the boundaries of the M2-branes and M5-branes. We call them ``M-strings'' \cite{Haghighat:2013gba}\cite{Hohenegger:2013ala}\cite{Haghighat:2013tka}\cite{Kim:2013nva}\cite{Kim:2014kta}\cite{Hohenegger:2015cba}\cite{Hayashi:2015fsa}. This configuration is dual to a $(p,q)$ 5-brane web \cite{Aharony:1997bh}. Furthermore, the $(p,q)$ 5-brane web is dual to the refined topological string \cite{Gopakumar:1998ii}\cite{Gopakumar:1998jq} on a non-compact toric Calabi-Yau manifold \cite{Leung:1997tw}\cite{Hollowood:2003cv}. Thus, we can calculate the partition function of M-strings by using the refined topological vertex formalism \cite{Aganagic:2002qg}\cite{Iqbal:2002we}\cite{Iqbal:2003ix}\cite{Iqbal:2003zz}\cite{Eguchi:2003sj}\cite{Aganagic:2003db}\cite{Awata:2005fa}\cite{Iqbal:2007ii}\cite{Taki:2007dh}\cite{Awata:2008ed}\cite{Iqbal:2012mt}.
\par

The M-strings have $\mathcal{N} = (4,4)$ supersymmetry in flat space \cite{Haghighat:2013gba}. When we put the M-strings on $A_{1}$ ALE space, the supersymmetry is broken to $\mathcal{N} = (4,0)$. Furthermore, in order to use the duality to the refined topological string, we consider some deformations by giving masses to the bifundamental hypermultiplets. These deformations do not break the $\mathcal{N} = (4,0)$ supersymmetry anymore \cite{Haghighat:2013tka}. According to \cite{Haghighat:2013gba}\cite{Hohenegger:2013ala}\cite{Haghighat:2013tka}, the partition function of M-strings agrees with the elliptic genus of the supersymmetric gauge theory through the chain of dualities. \par

In this paper, we consider one of the web diagrams which is dual to the above situation. This diagram is also related to the web diagram which is discussed in \cite{Haghighat:2013tka} under the flop transition \cite{Iqbal:2004ne}\cite{Konishi:2006ev}\cite{Taki:2008hb}. We calculate the partition function of M-strings by using the refined topological vertex formalism. Our result is consistent with \cite{Hohenegger:2013ala}\cite{Haghighat:2013tka} under the flop transition. 
%On the other hand, We calculate the elliptic genus of  $\mathcal{N}=(4,4)$ gauge theory by using the localization method\cite{Benini:2013nda}\cite{Benini:2013xpa}\cite{Gadde:2013dda}\cite{Harvey:2014nha}\cite{Honda:2015yha}. 
Then, we find that the supersymmetry of M-strings gets enhanced to $\mathcal{N}=(4,4)$ by tuning some K\"ahler parameters. Such a tuning of the K\"ahler parameters is discussed in the literature, for instance as in \cite{Hayashi:2013qwa}
 \par 

%However, we could not find the elliptic genus that is coincident with M-strings partition function completely.  This result about the elliptic genus depends on taking poles. Therefore, as a future work, we would like to consider this problem. Furthermore, we would like to consider why we obtain $\mathcal{N}=(4,4)$ structure. In general, the enhancement of supersymmetry does not occur because of $A_{1}$ ALE space. Thus we need to consider the effect more precisely when the K\"ahler parameters are tuned. In addition, in this paper we consider the simplest system. Then, as a straightforward extension, we can consider more general system. This also will have the structure of $\mathcal{N} = (4,4)$ supersymmetry. However, the reason why the partition function has $\mathcal{N}=(4,4)$ structure is due to the nontrivial cancellation of theta function. Therefore, this cancellation probably occurs if there are even number of M5-branes. Considering the meaning of this is also our future work.
%\par

The organization of this paper is as follows: In section \ref{sec:2}, we review two M5-branes on $A_{1}$ ALE space very briefly. We also introduce the web diagram which corresponds to the M2-M5 brane system. More detailed discussion about our model is written in \cite{Haghighat:2013gba}\cite{Hohenegger:2013ala}\cite{Haghighat:2013tka}. In section \ref{sec:3}, we calculate the partition function by using the refined topological vertex formalism. Then, by tuning some K\"ahler parameters, we find that the partition function of M-strings in $A_{1}$ ALE space agrees with that in flat space. This partition function is also coincident with the elliptic genus of the $\mathcal{N}=(4,4)$ $U(k)$ gauge theory for further setting some other K\"ahler parameters.  Finally, in section \ref{sec:4}, we discuss our results and another possibility of the enhancement of supersymmetry by tuning some K\"ahler parameters which are different from the above choices.

\section{Brane configuration and M-strings} 
\label{sec:2}
\begin{figure}[htbp]
    \begin{center}
    \includegraphics[clip,width=12cm]{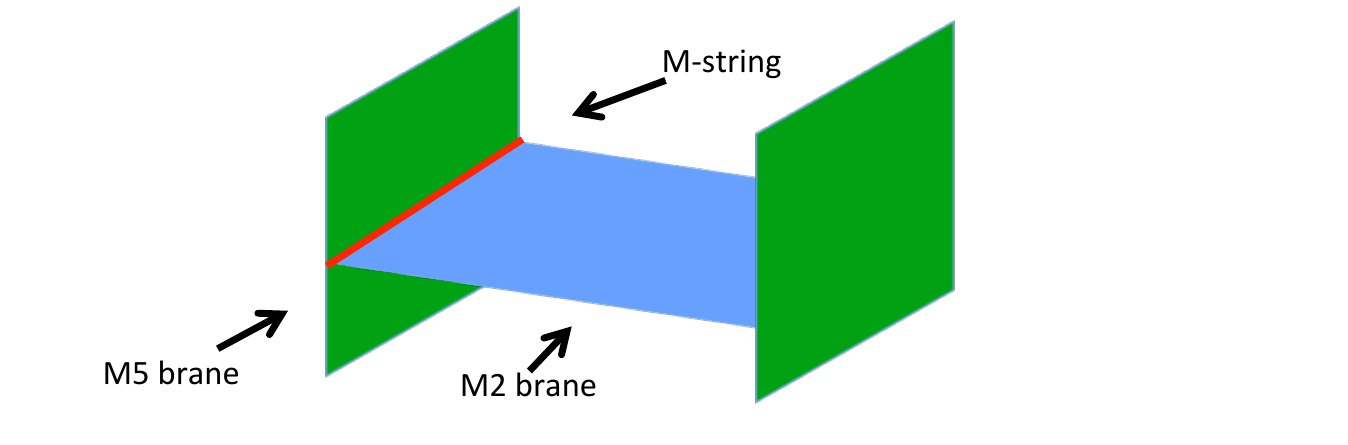}
    \caption{The brane configuration. The M2-brane, M5-branes, and M-strings are denoted by the blue sheet, green sheets, and red lines, respectively.}
    \label{system}
  \end{center}
\end{figure}

Consider the M2-M5 brane system on $A_{1}$ ALE space. We call the boundaries of M2- and M5-branes ``M-strings'' (see Fig. \ref{system}) \cite{Haghighat:2013gba}. According to a discussion in \cite{Haghighat:2013gba}, the M-strings have $\mathcal{N}=(4,4)$ supersymmetry in flat space. However the supersymmetry is broken to $\mathcal{N} = (4,0)$ on $A_{1}$ ALE space because of orbifolding \cite{Haghighat:2013tka}.
  \begin{table}[htb]
\begin{center}
  \begin{tabular}{|c||c|c|c|c|c|c|c|c|c|c|c|} \hline
     M theory & $X_{0}$ & $X_{1}$ & $X_{2}$ & $X_{3}$ & $X_{4}$ & $X_{5}$ & $X_{6}$ & $X_{7}$ 
     & $X_{8}$ & $X_{9}$ & $X_{\natural}$  \\ \hline \hline
    $2$ M5 & 
	$\circ$ & $\circ$ & $\circ$ & $\circ$ & $\circ$ & $\circ$  &  &  &  & & \\ \hline
    $k$ M2 & 
	$\circ$ & $\circ$ &  &  &  &  & $\circ$ &  &  & & \\ \hline
    $A_{1}$~ALE &  &  &  &  &  &  &  & $\circ$ & $\circ$ & $\circ$& $\circ$ \\ \hline
  \end{tabular}
  \caption{The brane setup where $A_{1}$ ALE space is spanned by $X_{7}$, $X_{8}$, $X_{9}$, $X_{\natural}$ and $k$ is the number of the M2-branes.}
    \end{center}
\end{table}
\par
\begin{figure}[htbp]
    \begin{center}
    \includegraphics[clip,width=12cm]{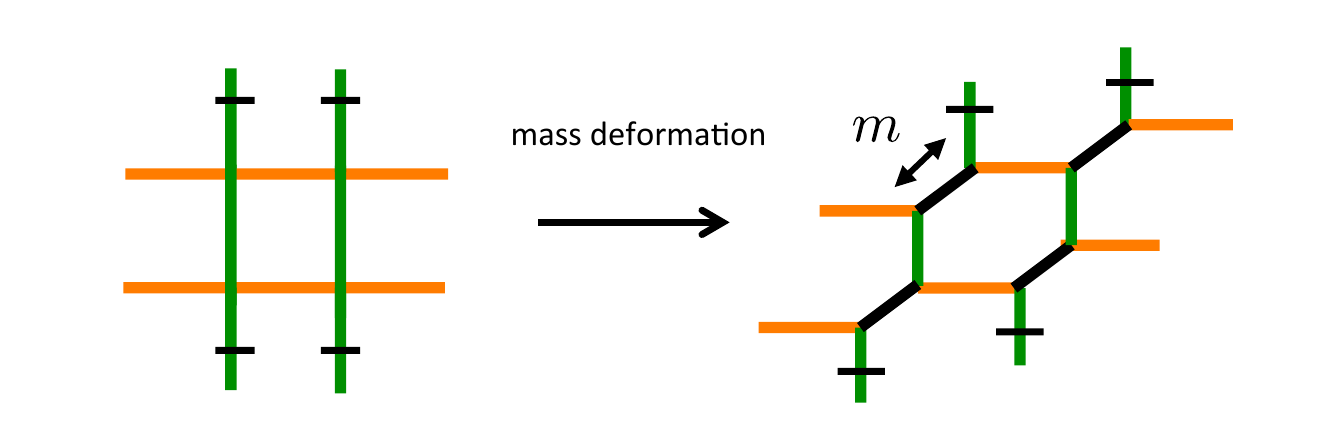}
    \caption{The brane web and its blow-up. The D5-branes and NS5-branes are denoted by the green lines and orange lines, respectively.}
    \label{blow-up}
  \end{center}
\end{figure}

 Let us consider the chain of dualities between M theory and Type IIB superstring theory \cite{Haghighat:2013tka}.  We set $X_{1}$ as M-theory circle. Then the M5-branes and the M2-branes become the D4-branes and the F1-branes, respectively. After T-duality along the $X_{7}$, we get the D5-NS5 brane web (see Fig. \ref{blow-up}). Then, in order to connect this geometry to the refined topological string, we introduce some mass parameters and $\Omega$-background by fibering over the circle.\footnote{Strictly speaking, we need consider a further compactification on another $S^1$, and we fiber the space over this circle. Its discussion is written in \cite{Haghighat:2013gba}\cite{Hohenegger:2013ala}\cite{Haghighat:2013tka}} These mass parameters correspond to the blow-up parameters in the web diagram. \par
\newpage
\par
\begin{figure}[htbp]
   \centering
    \includegraphics[width=12cm]{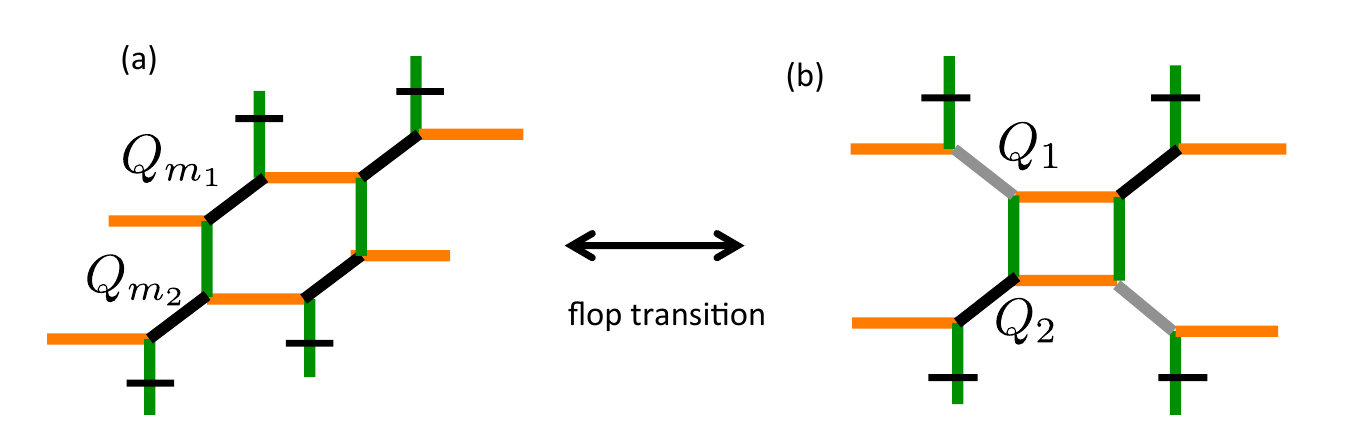}
    \caption{The flop transition. We define the K\"ahler parameters $Q_{m1}$, $Q_{m2}$, $Q_{1}$, and $Q_{2}$ which denote the mass deformations. The diagram (a) is discussed in \cite{Hohenegger:2013ala}\cite{Haghighat:2013tka}. We will discuss the diagram (b).}
            \label{conjecture}
\end{figure}

This web diagram is related to the web diagram in Fig. \ref{conjecture} (b) under the flop transition \cite{Taki:2008hb}. In section \ref{sec:3}, we will calculate the partition function of M-strings corresponding to the web diagram (b) by using the topological vertex formalism.

\section{Refined topological vertex}
\label{sec:3}
\subsection{Computation of partition function}
In section \ref{sec:2}, we have introduced the M-strings and discussed how to be related to the web diagram. In this section, we calculate the partition function of M-strings. 
\begin{figure}[htbp]
    \begin{center}
    \includegraphics[width=13cm]{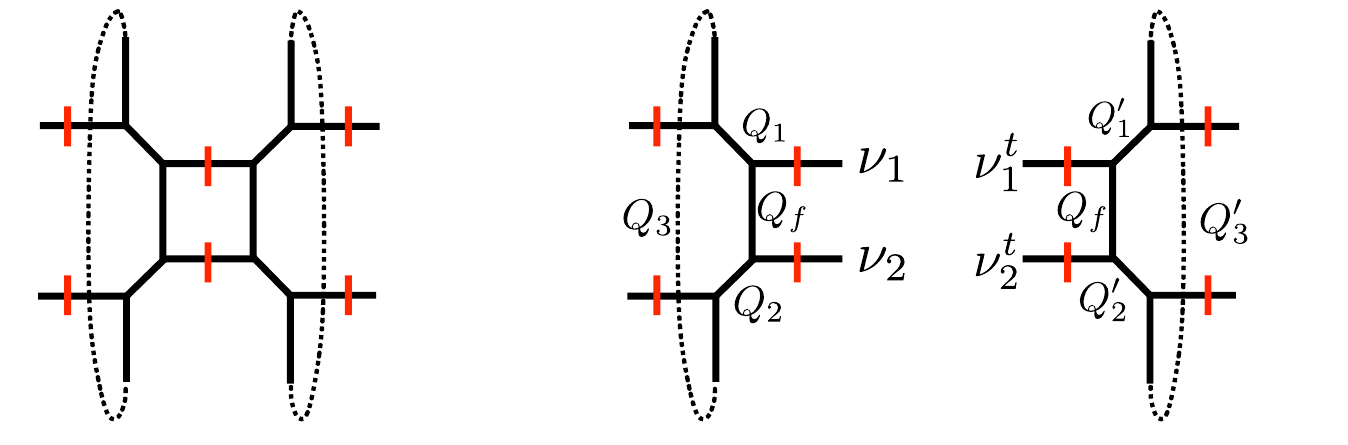}
    \caption{The figure (a) is the web diagram corresponding to the M2-M5 brane system. The parameters $Q_{3}$ and $Q_{f}$ denote the K\"ahler parameters along the vertical direction, whereas the parameter $Q_{b}$ denotes the K\"ahler parameter along the horizontal direction.  The figure (b1) and (b2) are its building blocks. The preferred direction is denoted by some red lines.}
    \label{web}
  \end{center}
\end{figure}\par
We define some K\"ahler parameters $Q_{1}$, $Q_{2}$, $Q_{3}$, ${Q'}_{1}$, ${Q'}_{2}$, ${Q'}_{3}$, $Q_{f}$, $Q_{b}$, and some Young diagrams $\nu_{1}$, $\nu_{2}$ as in Fig. \ref{web}, where $\nu_{1}^{t}$ and $\nu_{2}^{t}$ denote the transpose of $\nu_{1}$ and $\nu_{2}$ respectively. Then we write the partition function for the diagram (a) by using the refined topological vertex formalism,
\begin{eqnarray}
\label{partition function}
\mathcal{Z} = \sum_{\nu_{1},\nu_{2}}(-Q_{b})^{|\nu_{1}|+|\nu_{2}|}\mathcal{Z}_{\nu_{1}\nu_{2}}(t,q,Q)\mathcal{Z}_{\nu_{2}^{t}\nu_{1}^{t}}(q,t,Q')f_{\nu_{1}}(q,t)^{-1}f_{\nu_{2}}(q,t),
\end{eqnarray}
where the function $\mathcal{Z}_{\nu_{1}\nu_{2}}(t,q,Q)$ is 
the building block in Fig. \ref{web} (b1) which is defined by
\begin{eqnarray}
\mathcal{Z}_{\nu_{1}\nu_{2}}(t,q,Q) &:=& \sum_{\mu_{1},\mu_{2},\mu_{3},\mu_{f}}(-Q_{1})^{|\mu_{1}|}(-Q_{2})^{|\mu_{2}|}(-Q_{3})^{|\mu_{3}|}(-Q_{f})^{|\mu_{f}|} \nonumber \\ 
&&\times
C_{\mu_{1}^{t}\mu_{3}0}(t,q)C_{\mu_{1}\mu_{f}^{t}\nu_{1}}(q,t)
C_{\mu_{f}\mu_{2}^{t}\nu_{2}}(q,t)C_{\mu_{3}^{t}\mu_{2}0}(t,q)
\tilde{f}_{\mu_{f}^{t}}(q,t)\tilde{f}_{\mu_{3}^{t}}(q,t) ^{-1}.
\nonumber \\.
\label{block}
\end{eqnarray}
We also define the topological vertex $C_{\lambda \mu \nu}(t,q)$ and the framing factors $f_{\mu}(t,q)$ and $\tilde{f}_{\mu}(t,q)$ as
\begin{eqnarray}
C_{\lambda \mu \nu}(t,q) &=& t^{-\frac{||\mu^{t}||^{2}}{2}}q^{\frac{||\mu||^2 + ||\nu||^{2}}{2}} \tilde{Z}_{\nu}(t,q)\sum_{\eta}\Bigl(\frac{q}{t} \Bigr)^{\frac{|\eta| + |\lambda| - |\mu|}{2}}s_{\lambda^{t}/\eta}(t^{-\rho}q^{-\nu})s_{\mu/\eta}(t^{-\nu^{t}}q^{-\rho}),  
\label{def1}
\\ 
\tilde{Z}_{\nu}(t,q) &=& \prod_{(i,j) \in \nu}(1-q^{\nu_{i}-j}t^{\nu_{j}^{t} -i +1})^{-1},\\
f_{\mu} (t,q) &=& (-1)^{|\mu|}
q^{-\frac{||\mu||^2}{2}}t^{\frac{||\mu^{t}||^2}{2}},\\
\tilde{f}_{\mu} (t,q) &=& (-1)^{|\mu|}
\Bigl(\frac{t}{q}\Bigr)^{\frac{|\mu|}{2}}
q^{-\frac{||\mu||^2}{2}}t^{\frac{||\mu^{t}||^2}{2}}
.
\label{def4}
\end{eqnarray}
Physically, the variables $t$ and $q$ are related to the gauge theory parameters (or $\Omega$-background) as follows,
\begin{eqnarray}
 t = \mathrm{e}^{-2 \pi i \epsilon_{1}},~~q= \mathrm{e}^{2 \pi i \epsilon_{2}}.
\end{eqnarray}
By substituting above definitions into (\ref{block}), we obtain
\begin{eqnarray}
\mathcal{Z}_{\nu_{1}\nu_{2}}(t,q,Q) &=& \sum_{\{\mu\},\{\eta\}} (-Q_{1})^{|\mu_{1}|}(-Q_{2})^{|\mu_{2}|}Q_{3}^{|\mu_{3}|}Q_{f}^{|\mu_{f}|}
\nonumber \\ && \times
\tilde{Z}_{\nu_{1}}(q,t)\tilde{Z}_{\nu_{2}}(q,t)
t^{\frac{||\nu_{1}||^2 + ||\nu_{2}||^2}{2}}\Bigl(\frac{q}{t}\Bigr)^{\frac{|\eta_{1}|-|\eta_{2}|-|\eta_{3}|+|\eta_{4}|}{2}}\Bigl(\frac{t}{q}\Bigr)^{\frac{|\mu_{3}|}{2}
-\frac{|\mu_{f}|}{2}}
\nonumber \\ && \times
s_{\mu_{1}/\eta_{1}}(t^{-\rho})s_{\mu_{3}/\eta_{1}}(q^{-\rho})
s_{\mu_{1}^{t}/\eta_{2}}(q^{-\rho} t^{-\nu_{1}^{t}})s_{\mu_{f}^{t}/\eta_{2}}(t^{-\rho} q^{-\nu_{1}})
\nonumber  \\ && \times
s_{\mu_{f}^{t}/\eta_{3}}(q^{-\rho} t^{-\nu_{2}})s_{\mu_{2}^{t}/\eta_{3}}(q^{-\nu_{2}^{t}}t^{-\rho})
s_{\mu_{3}/\eta_{4}}(t^{-\rho})s_{\mu_{2}/\eta_{4}}(q^{-\rho})
.
\label{build.block}
\end{eqnarray}
We can calculate (\ref{build.block}) by using some formulas in Appendix. Then we obtain
\begin{eqnarray}
\hat{\mathcal{Z}}_{\nu_{1}\nu_{2}}(t,q,Q) &=& \tilde{Z}_{\nu_{1}}(q,t)\tilde{Z}_{\nu_{2}}(q,t)t^{\frac{1}{2}(||\nu_{1}||^2+||\nu_{2}||^2)} 
\nonumber \\ &&\times
\prod_{i,j=1}^{\infty}\prod_{n=0}^{\infty}\Biggl[ \Biggl\{
\frac{1-Q_{1}\Lambda^{n}t^{-\nu_{1,i}+j-\frac{1}{2}}q^{i-\frac{1}{2}}}{1-Q_{1}\Lambda^{n}t^{j-\frac{1}{2}}q^{i-\frac{1}{2}}}
\frac{1-Q_{1}^{-1}\Lambda^{n+1}t^{j-\frac{1}{2}}q^{-\nu^{t}_{1,j}+i-\frac{1}{2}}}{1-Q_{1}^{-1}\Lambda^{n+1}t^{j-\frac{1}{2}}q^{i-\frac{1}{2}}} 
\nonumber \\ &&\times
\frac{1-Q_{2}\Lambda^{n}t^{j-\frac{1}{2}}q^{-\nu_{2,j}^{t}+i-\frac{1}{2}}}{1-Q_{2}\Lambda^{n}t^{j-\frac{1}{2}}q^{i-\frac{1}{2}}}
\frac{1-Q_{2}^{-1}\Lambda^{n+1}t^{-\nu_{2,i}+j-\frac{1}{2}}q^{i-\frac{1}{2}}}{1-Q_{2}^{-1}\Lambda^{n+1}t^{j-\frac{1}{2}}q^{i-\frac{1}{2}}} 
\nonumber \\ &&\times
\frac{1-Q_{1}Q_{3}\Lambda^{n}t^{-\nu_{1,i}+j-\frac{1}{2}}q^{i-\frac{1}{2}}}{1-Q_{1}Q_{3}\Lambda^{n}t^{j-\frac{1}{2}}q^{i-\frac{1}{2}}}
\frac{1-(Q_{1}Q_{3})^{-1}\Lambda^{n+1}t^{i-\frac{1}{2}}q^{-\nu^{t}_{1,j}+j-\frac{1}{2}}}{1-(Q_{1}Q_{3})^{-1}\Lambda^{n+1}t^{i-\frac{1}{2}}q^{j-\frac{1}{2}}}
\nonumber \\ &&\times
\frac{1-Q_{2}Q_{3}\Lambda^{n}t^{j-\frac{1}{2}}q^{-\nu_{2,j}^{t}+i-\frac{1}{2}}}{1-Q_{2}Q_{3}\Lambda^{n}t^{j-\frac{1}{2}}q^{i-\frac{1}{2}}}
\frac{1-(Q_{2}Q_{3})^{-1}\Lambda^{n+1}t^{-\nu_{2,i}+j-\frac{1}{2}}q^{i-\frac{1}{2}}}{1-(Q_{2}Q_{3})^{-1}\Lambda^{n+1}t^{j-\frac{1}{2}}q^{i-\frac{1}{2}}}
\Biggr\} 
\nonumber \\ &&\times
\Biggl\{
\frac{1-Q_{f}\Lambda^{n}t^{-\nu_{2,i}+j-1}q^{-\nu^{t}_{1,j}+i}}{1-Q_{f}\Lambda^{n}t^{j-1}q^{i}}
\frac{1-Q_{f}^{-1}\Lambda^{n+1}t^{-\nu_{1,i}+j-1}q^{-\nu^{t}_{2,j}+i}}{1-Q_{f}^{-1}\Lambda^{n+1}t^{i-1}q^{j}}
\nonumber \\ &&\times
\frac{1-\Lambda^{n+1}t^{-\nu_{1,i}+j-1}q^{-\nu^{t}_{1,j}+i}}{1-\Lambda^{n+1}t^{j-1}q^{i}}
\frac{1-\Lambda^{n+1}t^{-\nu_{2,i}+j-1}q^{-\nu_{2,j}^{t}+i}}{1-\Lambda^{n+1}t^{j-1}q^{i}}
\Biggr\}^{-1} \Biggr],\nonumber \\
(\Lambda &:=& Q_{1}Q_{2}Q_{3}Q_{f}),
\nonumber \\ 
\label{1st building block}
\end{eqnarray}
where we divide (\ref{1st building block}) by the trivial building block,
\begin{eqnarray}
\hat{\mathcal{Z}}_{\nu_{1}\nu_{2}} := \frac{\mathcal{Z}_{\nu_{1}\nu_{2}}}{\mathcal{Z}_{\emptyset \emptyset}}.
\end{eqnarray}
After some calculation, we obtain the building block in Fig. \ref{web} (b1),
\begin{eqnarray}
\hat{\mathcal{Z}}_{\nu_{1}\nu_{2}}(t,q,Q)& = &
\tilde{Z}_{\nu_{1}}(q,t)\tilde{Z}_{\nu_{2}}(q,t)t^{\frac{1}{2}(||\nu_{1}||^2+||\nu_{2}||^2)} 
 \nonumber \\ &&\times
\prod_{n=0}^{\infty}
\prod_{(i,j) \in \nu_{1}} 
\Biggl \{
\frac{(1-Q_{1}\Lambda^{n}t^{-\nu_{1,i}+j-\frac{1}{2}}q^{i-\frac{1}{2}})(1-Q_{1}^{-1}\Lambda^{n+1}t^{\nu_{1,i}-j+\frac{1}{2}}q^{-i+\frac{1}{2}})}{(1-Q_{f}\Lambda^{n}t^{\nu_{1,i}-j}q^{\nu^{t}_{2,j}-i+1})(1-Q_{f}^{-1}\Lambda^{n+1}t^{-\nu_{1,i}+j-1}q^{-\nu_{2,j}^{t} +i})}
\nonumber \\ &&\times
\frac{
(1-Q_{1}Q_{3}\Lambda^{n}t^{-\nu_{1,i}+j-\frac{1}{2}}q^{i-\frac{1}{2}})
(1-Q_{1}^{-1}Q_{3}^{-1}\Lambda^{n+1}t^{\nu_{1,i}-j+\frac{1}{2}}q^{-i+\frac{1}{2}})
}{
(1-\Lambda^{n+1}t^{-\nu_{1,i}+j-1}q^{-\nu^{t}_{1,j}+i})
(1-\Lambda^{n+1}t^{\nu_{1,i}-j}q^{\nu^{t}_{1,j}-i+1})
}
 \Biggr \}
\nonumber \\ &&\times
\prod_{(i,j) \in \nu_{2}} 
\Biggl \{
\frac{(1-Q_{2}\Lambda^{n}t^{\nu_{2,i}-j+\frac{1}{2}}q^{-i+\frac{1}{2}})(1-Q_{2}^{-1}\Lambda^{n+1}t^{-\nu_{2,i}+j-\frac{1}{2}}q^{i-\frac{1}{2}})}{(1-Q_{f}\Lambda^{n}t^{-\nu_{2,i}+j-1}q^{-\nu^{t}_{1,j}+i})(1-Q_{f}^{-1}\Lambda^{n+1}t^{\nu_{2,i}-j}q^{\nu^{t}_{1,j}-i+1})} 
\nonumber \\ &&\times
\frac{
(1-Q_{2}Q_{3}\Lambda^{n}t^{\nu_{2,i}-j+\frac{1}{2}}q^{-i+\frac{1}{2}})
(1-Q_{2}^{-1}Q_{3}^{-1}\Lambda^{n+1}t^{-\nu_{2,i}+j-\frac{1}{2}}q^{i-\frac{1}{2}})
}{
(1-\Lambda^{n+1}t^{-\nu_{2,i}+j-1}q^{-\nu_{2,j}^{t}+i})
(1-\Lambda^{n+1}t^{\nu_{2,i}-j}q^{\nu_{2,j}^{t}-i+1})
}
 \Biggr \},
\nonumber \\
\label{result1}
\end{eqnarray}
and the building block in Fig. \ref{web} (b2),
\begin{eqnarray}
\hat{\mathcal{Z}}_{\nu^{t}_{2}\nu^{t}_{1}}(q,t,Q')& = &\tilde{Z}_{\nu^{t}_{1}}(t,q)\tilde{Z}_{\nu^{t}_{2}}(t,q)q^{\frac{1}{2}(||\nu_{1}^{t}||^2+||\nu_{2}^{t}||^2)} 
 \nonumber \\ &&\times
\prod_{n=0}^{\infty}
\prod_{(i,j) \in \nu_{1}} 
 \Biggl \{
\frac{(1-Q'_{1}\Lambda^{n}t^{-\nu_{1,i}+j-\frac{1}{2}}q^{i-\frac{1}{2}})(1-{Q'}_{1}^{-1}\Lambda^{n+1}t^{\nu_{1,i}-j+\frac{1}{2}}q^{-i+\frac{1}{2}})
}{
(1-Q_{f}\Lambda^{n}t^{\nu_{1,i}-j+1}q^{\nu_{2,j}^{t}-i})
(1-Q_{f}^{-1}\Lambda^{n+1}t^{-\nu_{1,i}+j}q^{-\nu^{t}_{2,j}+i-1})
}
\nonumber \\ &&\times
\frac{
(1-{Q'}_{1}{Q'}_{3}\Lambda^{n}t^{-\nu_{1,i}+j-\frac{1}{2}}q^{i-\frac{1}{2}})
(1-{Q'}_{1}^{-1}{Q'}_{3}^{-1}\Lambda^{n+1}t^{\nu_{1,i}-j+\frac{1}{2}}q^{-i+\frac{1}{2}})
}{
(1-\Lambda^{n+1}t^{-\nu_{1,i}+j}q^{-\nu^{t}_{1,j}+i-1})
(1-\Lambda^{n+1}t^{\nu_{1,i}-j+1}q^{\nu^{t}_{1,j}-i})
}
 \Biggr \}
\nonumber \\ &&\times
\prod_{(i,j) \in \nu_{2}} 
\Biggl \{
\frac{(1-{Q'}_{2}\Lambda^{n}t^{\nu_{2,i}-j+\frac{1}{2}}q^{-i+\frac{1}{2}})(1-{Q'}_{2}^{-1}\Lambda^{n+1}t^{-\nu_{2,i}+j-\frac{1}{2}}q^{i-\frac{1}{2}})
}{
(1-Q_{f}\Lambda^{n}t^{-\nu_{2,i}+j}q^{-\nu^{t}_{1,j}+i-1})
(1-Q_{f}^{-1}\Lambda^{n+1}t^{\nu_{2,i}-j+1}q^{\nu^{t}_{1,j}-i})
} 
\nonumber \\ &&\times
\frac{
(1-{Q'}_{2}{Q'}_{3}\Lambda^{n}t^{\nu_{2,i}-j+\frac{1}{2}}q^{-i+\frac{1}{2}})
(1-{Q'}_{2}^{-1}{Q'}_{3}^{-1}\Lambda^{n+1}t^{-\nu_{2,i}+j-\frac{1}{2}}q^{i-\frac{1}{2}})
}{
(1-\Lambda^{n+1}t^{-\nu_{2,i}+j}q^{-\nu_{2,j}^{t}+i-1})
(1-\Lambda^{n+1}t^{\nu_{2,i}-j+1}q^{\nu_{2,j}^{t}-i})
}
 \Biggr \}.
\nonumber \\
\label{result2}
\end{eqnarray}
By substituting (\ref{result1}), (\ref{result2}) into (\ref{partition function}), we find
\begin{eqnarray}
\hat{\mathcal{Z}} &=& \sum_{\nu_{1},\nu_{2}} 
\Bigl(-\bar{Q}_{b}\sqrt{\frac{t}{q}}\Bigr)^{|\nu_{1}|} \Bigl(-\bar{Q'}_{b}\sqrt{\frac{t}{q}}\Bigr)^{|\nu_{2}|}
 \nonumber \\ &&\times
 \prod_{(i,j) \in \nu_{1}}
 \frac{
 \theta_{1}(\tau ; u^{1}_{ij})\theta_{1}(\tau ; {U}^{2f}_{ij})\theta_{1}(\tau ; u'^{1}_{ij})\theta_{1}(\tau ; {u'}^{13}_{ij})
 }{
 \theta_{1}(\tau ; y^{123}_{ij})\theta_{1}(\tau ; v^{\nu_{1}}_{ij})\theta_{1}(\tau ; Y^{f}_{ij})\theta_{1}(\tau ; V^{\nu_{1}}_{ij})
 } 
 \nonumber \\ &&\times
\prod_{(i,j) \in \nu_{2}}\frac{\theta_{1}(\tau ; \tilde{u}'^{2}_{ij})\theta_{1}(\tau ; \tilde{U}'^{1f}_{ij})\theta_{1}(\tau ; \tilde{u}^{2}_{ij})\theta_{1}(\tau ; \tilde{u}^{23}_{ij})}{\theta_{1}(\tau ; w^{f}_{ij})\theta_{1}(\tau ; v^{\nu_{2}}_{ij})\theta_{1}(\tau ; W^{123}_{ij})\theta_{1}(\tau ; V^{\nu_{2}}_{ij})},
\label{partition function theta}
\end{eqnarray}
where we define some variables as follows:
\begin{eqnarray}
\mathrm{e}^{2 \pi i \tau} &=& \Lambda,
~~ \bar{Q}_{b} = Q_{b}\sqrt{ Q_{1}Q'_{1}Q_{2}/Q'_{2}},
~~ \bar{Q'}_{b} = Q_{b}\sqrt{ Q'_{1}Q'_{2}Q_{2}/Q_{1}},
\nonumber \\
\mathrm{e}^{2 \pi i u^{I}_{ij}} &=& Q_{I}^{-1} t^{\nu_{1,i}-j+\frac{1}{2}}q^{-i+\frac{1}{2}},~~
\mathrm{e}^{2 \pi i \tilde{u}^{I}_{ij}} = Q_{I}^{-1} t^{-\nu_{2,i}+j-\frac{1}{2}}q^{i-\frac{1}{2}},
\nonumber \\
\mathrm{e}^{2 \pi i U^{I}_{ij}} &=& Q_{I}^{-1}t^{-\nu_{1,i}+j-\frac{1}{2}}q^{i-\frac{1}{2}},~~
\mathrm{e}^{2 \pi i \tilde{U}^{I}_{ij}} = Q_{I}^{-1}t^{\nu_{2,i}-j+\frac{1}{2}}q^{-i+\frac{1}{2}},
\nonumber \\
\mathrm{e}^{2 \pi i y^{I}_{ij}} &=& Q_{I}^{-1}t^{\nu_{1,i}-j+1}q^{\nu^{t}_{2,j}-i},~~
\mathrm{e}^{2 \pi i Y^{I}_{ij}} = Q_{I}^{-1}t^{-\nu_{1,i}+j}q^{-\nu^{t}_{2,j}+i-1},
 \nonumber \\
\mathrm{e}^{2 \pi i w^{I}_{ij}} &=& Q_{I}^{-1}t^{\nu_{2,i}-j+1}q^{\nu^{t}_{1,j}-i},~~
\mathrm{e}^{2 \pi i W^{I}_{ij}} = Q_{I}^{-1}t^{-\nu_{2,i}+j}q^{-\nu^{t}_{1,j}+i-1},
 \nonumber \\
\mathrm{e}^{2 \pi i v^{\nu_{I}}_{ij}} &=& t^{\nu_{I,i}-j+1}q^{\nu_{I,j}^{t}-i},~~
\mathrm{e}^{2 \pi i V^{\nu_{I}}_{ij}} = t^{\nu_{I,i}-j}q^{\nu_{I,j}^{t}-i+1}.
\label{partition function result}
\end{eqnarray}
This expression is slightly different from \cite{Haghighat:2013tka}. However one can show that (\ref{partition function result}) agrees with the result in \cite{Haghighat:2013tka} under the flop transition \cite{Taki:2008hb}
\footnote{Where we impose the condition $\frac{Q_{2}}{Q_{1}} = \frac{Q'_{1}}{Q'_{2}}$ which is also used in the reference \cite{Haghighat:2013tka}}.

\subsection{Enhancement of Supersymmetry}
In this subsection, we observe the enhancement of supersymmetry by tuning the K\"ahler parameters to special values. Consider the following constraints,
\begin{eqnarray}
 Q_{2}=\sqrt{\frac{q}{t}} ,Q'_{2}=\sqrt{\frac{t}{q}}(=Q_{2}^{-1}).
\label{tune}
\end{eqnarray}
Then, the infinite products $\prod_{n=0}^{\infty}\prod_{(i,j) \in \nu_{2}} (1-\Lambda^{n}t^{-\nu_{2,i}+j}q^{i-1})$ in (\ref{result2}) become zero unless the Young diagram $\nu_{2}$ becomes empty. Thus, after some cancellations, we find
\begin{eqnarray}
\hat{\mathcal{Z}} =
 \sum_{\nu_{1}}(-Q_{b}\sqrt{Q_{1}Q'_{1}})^{|\nu_{1}|}
 \prod_{(i,j) \in \nu_{1}}
 \frac
 {
 \theta_{1}(\tau ;u_{ij}^{1} )
 \theta_{1}(\tau ; u_{ij}^{-1'})
  }{
  \theta_{1}(\tau ;v^{\nu_{1}}_{ij} )
  \theta_{1}(\tau ; V^{\nu_{1}}_{ij})
  },
\end{eqnarray}
where we define
\begin{eqnarray}
\mathrm{e}^{2 \pi i u^{-I}_{ij}} = Q_{I}t^{-\nu_{1,i}+j-\frac{1}{2}}q^{i-\frac{1}{2}}.
\end{eqnarray}
Therefore, we obtain the partition function of $k$ M-strings as follows,
\begin{eqnarray}
\mathcal{Z}^{k}_{M-strings} &=& \sum_{|\nu_{1}|=k}
 \prod_{(i,j) \in \nu_{1}}
 \frac
 {
 \theta_{1}(\tau ;u_{ij}^{1} )
 \theta_{1}(\tau ; u_{ij}^{-1})
  }{
  \theta_{1}(\tau ;v^{\nu_{1}}_{ij} )
  \theta_{1}(\tau ; V^{\nu_{1}}_{ij})
  }.
\label{k-string}
\end{eqnarray}
\par
This partition function agrees with the partition function of M-strings in flat space when we tune the K\"ahler parameters as follow,
\begin{eqnarray}
Q_{1} ={Q'}^{-1}_{1}.
\end{eqnarray}
 According to a discussion in \cite{Haghighat:2013gba}, by tuning the K\"ahler parameters as follows,
\begin{eqnarray}
Q_{1} = {Q'}^{-1}_{1}=\sqrt{\frac{t}{q}},
\end{eqnarray}
and performing appropriate normalization, the partition function of M-strings (\ref{k-string}) agrees with the elliptic genus of the $\mathcal{N}=(4,4)$ $U(k)$ pure gauge theory.
\par
\begin{figure}[htbp]
    \begin{center}
    \includegraphics[clip,width=12cm]{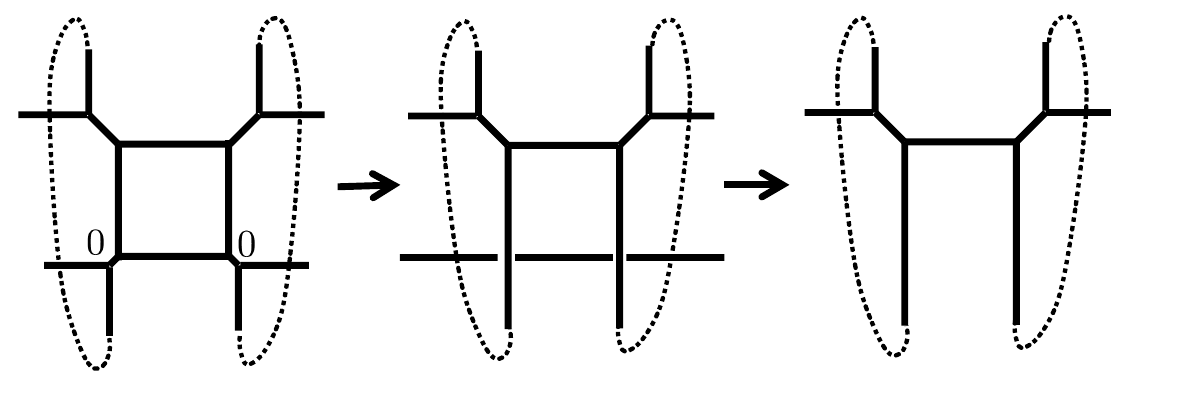}
    \caption{The geometric transition. We can remove the framing and the result is on the right figure.}
    \label{geometric}
  \end{center}
\end{figure}
\begin{figure}[htbp]
    \begin{center}
    \includegraphics[clip,width=12cm]{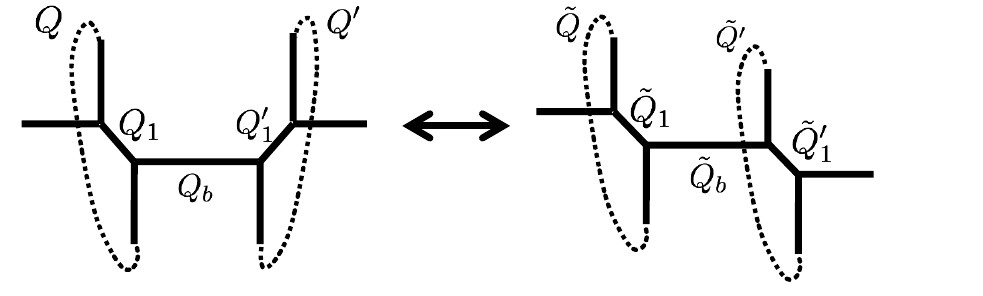}
    \caption{The flop transition.}
    \label{flop}
  \end{center}
\end{figure}

We interpret this result by using the geometric transition \cite{Gopakumar:1998ii}\cite{Taki:2010bj}. In unrefined case ($t=q$), the K\"ahler parameters become 1,
\begin{eqnarray}
Q_{2} =Q'_{2} =1.
\end{eqnarray}
Usually, by the geometric transition\cite{Gopakumar:1998ii}, the K\"ahler parameters become multiplicity of branes. However, in this case, the K\"ahler parameters are 1. This means that there are no branes. Thus, we can remove the framing as in Fig. \ref{geometric}.
\\
The right figure of Fig. \ref{geometric} is related to the web diagram which is considered in \cite{Haghighat:2013gba}\cite{Haghighat:2013tka} under the flop transition (see Fig. \ref{flop}). 
The K\"ahler parameters are related as follows,
\begin{eqnarray}
&&Q'=\tilde{Q'}_{1}\tilde{Q'}\tilde{Q'}_{1},~~Q'_{1}=\tilde{Q'}^{-1}_{1},~~Q_{b}=\tilde{Q'}_{1}\tilde{Q}_{b}.
\end{eqnarray}
\par
Thus, we observe that the supersymmetry of M-strings gets enhanced to the $\mathcal{N}=(4,4)$ by tuning the K\"ahler parameters.
\par
Moreover, we can interpret this enhancement in terms of the 2 dimensional field theory which is dual to the M-strings.
\par
According to a discussion in \cite{Haghighat:2013tka}\cite{Okuyama:2005gq}, this M2-M5 brane system is dual to the D1-D5 brane system on $A_{1}$ ALE space. $k$ M2-branes, $2$ M5-branes, and $A_{1}$ ALE space  correspond to $k$ D1-branes, $A_{1}$ ALE space, and 2 D5-branes, respectively. 
  \begin{table}[htb]
\begin{center}
  \begin{tabular}{|c||c|c|c|c|c|c|c|c|c|c|} \hline
     IIB theory & $X_{0}$ & $X_{1}$ & $X_{2}$ & $X_{3}$ & $X_{4}$ & $X_{5}$ & $X_{6}$ & $X_{7}$ 
     & $X_{8}$ & $X_{9}$   \\ \hline \hline
    $2$ D5 & 
	$\circ$ & $\circ$ & $\circ$ & $\circ$ & $\circ$ & $\circ$  &  &  &  &  \\ \hline
    $k$ D1 & 
	$\circ$ & $\circ$ &  &  &  &  & &  &  &  \\ \hline
    $A_{1}$~ALE &  &  &  &  &  &  & $\circ$  & $\circ$ & $\circ$ & $\circ$  \\ \hline
  \end{tabular}
  \caption{The brane setup which is dual to the M2-M5 brane system in section \ref{sec:2}. The M-strings correspond to $k$ D1-branes.}
    \end{center}
\end{table}\\
In this case, the theory on $k$ D1-branes is the $\mathcal{N}=(4,0)$ $U(k)$ gauge theory with one adjoint, two chiral multiplets and two fermi multiplets. If we remove the D5-branes, the theory on $k$ D1-branes is the $\mathcal{N}=(4,4)$ $U(k)$ pure gauge theory. Therefore, in terms of duality, removal of the framing corresponds to removal of the D5-branes.
\par
\begin{figure}[htbp]
    \centering
    \includegraphics[width=16cm]{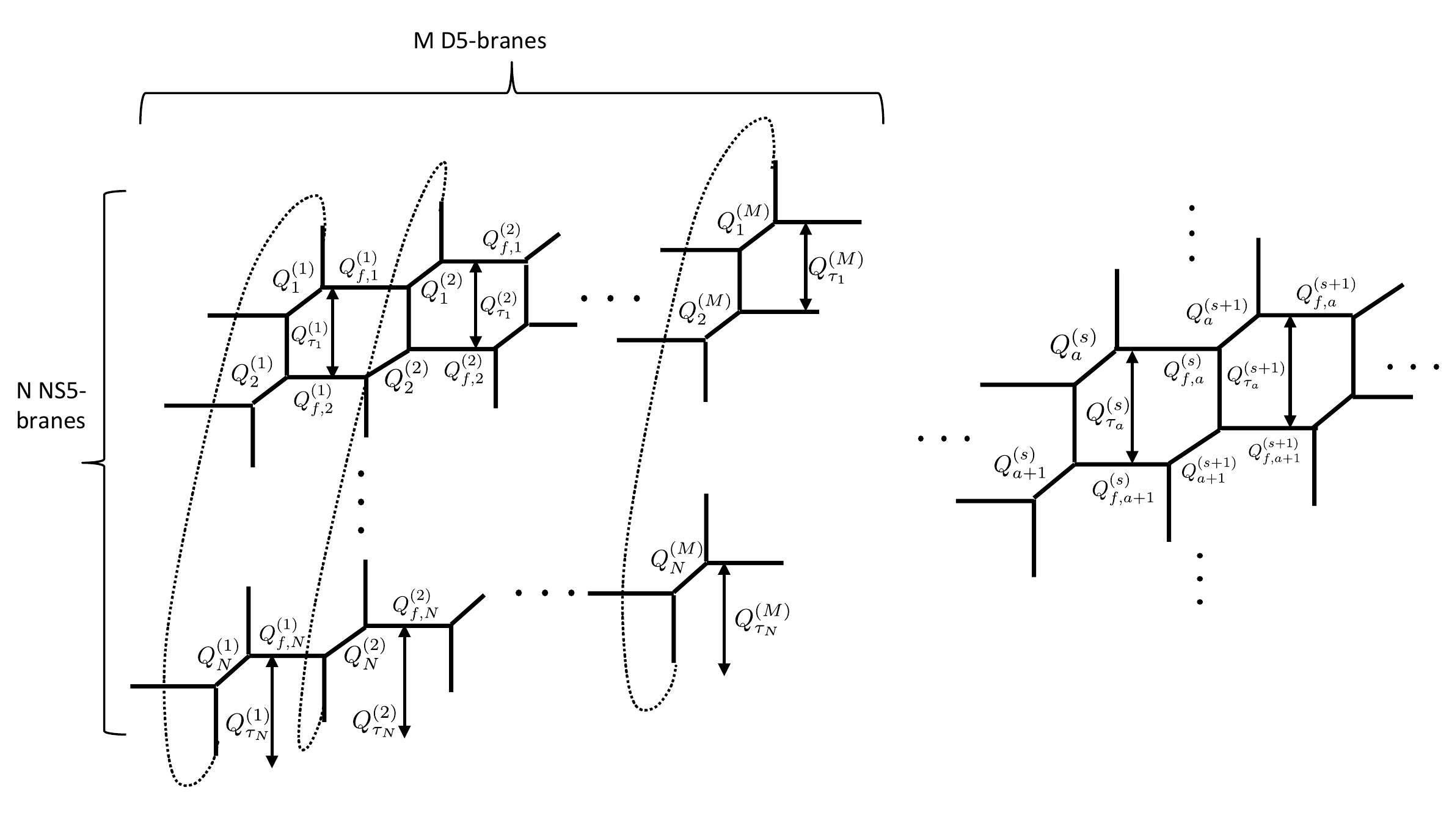}
    \caption{The type IIB brane web which is dual to the M theory system. The right figure is the fraction of the web diagram. The K\"ahler parameters $Q_{a}^{(s)}$ and $Q_{\tau_{a}}^{(s)}$ denote the mass deformations and the distance between two NS5-branes, respectively. The parameters  $Q_{f,a}^{(s)}$ denote the K\"ahler parameters along the horizontal direction.}
    \label{general web}
    %\ref{general web}
\end{figure}

We can generalize this enhancement to the theory on multiple M5-branes on $A_{N-1}$ ALE space. This M theory system is dual to the $(p,q)$ 5-brane web which is drawn in Fig. \ref{general web}. By using the result in \cite{Haghighat:2013tka}, the partition function of $M$ M5 branes on $A_{N-1}$ singularity is as follow,
\begin{eqnarray}
\hat{\mathcal{Z}}_{M}^{A_{N-1}} =
\sum_{\text{all indices}}\prod_{s=1}^{M-1}\prod_{a=1}^{N}
\Bigl(\bar{Q}_{f,a}^{(s)}\Bigr)^{|\mu_{a}^{(s)}|}
\prod_{(i,j) \in \mu_{a}^{(s)}}\prod_{b=1}^{N}
\frac{
\theta_{1}(\tau;z_{ab}^{(s)}(i,j))
\theta_{1}(\tau;w_{ab}^{(s)}(i,j))
}{
\theta_{1}(\tau;u_{ab}^{(s)}(i,j))
\theta_{1}(\tau;v_{ab}^{(s)}(i,j))
},
\end{eqnarray}
where we define
\begin{eqnarray}
\mathrm{e}^{2\pi z_{ab}^{(s)}(i,j)}
&=&
\Bigl(Q_{ab}^{(s+1)}\Bigr)^{-1}
t^{-\mu_{a,i}^{(s)}+j-1/2}q^{-\mu_{b,j}^{(s+1),t}+i-1/2},
\nonumber \\
\mathrm{e}^{2\pi w_{ab}^{(s)}(i,j)}
&=&
\Bigl(Q_{ba}^{(s)}\Bigr)^{-1}
t^{\mu_{a,i}^{(s)}-j+1/2}q^{\mu_{b,j}^{(s-1),t}-i+1/2},
\nonumber \\
\mathrm{e}^{2\pi u_{ab}^{(s)}(i,j)}
&=&
\Bigl(\hat{Q}_{ba}^{(s)}\Bigr)^{-1}
t^{\mu_{a,i}^{(s)}-j}q^{\mu_{b,j}^{(s),t}-i+1},
\nonumber \\
\mathrm{e}^{2\pi v_{ab}^{(s)}(i,j)}
&=&
\Bigl(\hat{Q}_{ab}^{(s)}\Bigr)^{-1}
t^{-\mu_{a,i}^{(s)}+j-1}q^{-\mu_{b,j}^{(s),t}+i},
\end{eqnarray}
and the some parameters $\bar{Q}_{f,a}^{(s)}$, $Q_{ab}^{(s)}$, and $\hat{Q}_{ab}^{(s)}$ are defined as follows,
\begin{eqnarray}
\bar{Q}_{f,a}^{(s)} &=& \Bigl( \frac{q}{t} \Bigr)^{\frac{N-1}{2}} Q_{f,a}^{(s)}\prod_{b=1}^{N}Q_{b}^{(s)},
\nonumber \\
\hat{Q}_{ab}^{(s)}
&=&
\begin{cases}
    1,~~~~\text{for}~a=b \\
    \tilde{Q}_{ab}^{(s)},~~~~\text{for}~a\neq b
\end{cases}
\nonumber \\
\tilde{Q}_{ab}^{(s)}
&=&
\begin{cases}
    \prod_{i=b}^{a-1}Q_{\tau_{i}}^{(s)},~~~~\text{for}~a>b \\
    Q_{\tau},~~~~\text{for}~a=b \\
    Q_{\tau}/\prod_{i=a}^{b-1}Q^{(s)}_{\tau_{i}},~~~~\text{for}~a<b
\end{cases}
\nonumber \\
Q_{ab}^{(s)}
&=&
\begin{cases}
    Q_{a}^{(s)}\prod_{i=b}^{N}Q_{\tau_{i}}^{(s)},~~~~(\text{mod}~Q_{\tau})~~~~\text{for}~a=1 \\
    Q_{a}^{(s)}\prod_{i=1}^{a-1}Q_{\tau_{i}}^{(s)}\prod_{j=b}^{N}Q_{\tau_{j}}^{(s)}.
    ~~~~(\text{mod}~Q_{\tau})~~~~\text{for}~a\neq 1
\end{cases}
\label{def}
%(\ref{def})
\end{eqnarray}
\\
If we tune the K\"ahler parameters at the fixed $a$,
\begin{eqnarray}
Q_{a}^{(s)}=\sqrt{\frac{t}{q}},~~~s=1,2,...,M
\label{setting the Kahler parameters}
%(\ref{setting the Kahler parameters})
\end{eqnarray}
the contribution of $\mu_{a}^{(s)}$,
\begin{eqnarray}
\prod_{s=1}^{M-1}
\prod_{(i,j) \in \mu_{a}^{(s)}}\prod_{b=1}^{N}
\frac{
\theta_{1}(\tau;z_{ab}^{(s)}(i,j))
\theta_{1}(\tau;w_{ab}^{(s)}(i,j))
}{
\theta_{1}(\tau;u_{ab}^{(s)}(i,j))
\theta_{1}(\tau;v_{ab}^{(s)}(i,j))
},
\end{eqnarray}
becomes zero unless the Young diagram $\mu_{a}^{(s)}$ becomes empty.  After several cancellation, the partition function becomes $\hat{\mathcal{Z}}_{M}^{A_{N-2}}$. In terms of this web diagram, this corresponds to remove the $a$-th framing by using the geometric transition. In terms of the D1-D5 system on $A_{M-1}$ ALE space, this corresponds to remove one of $N$ D5-branes.
\par
For example, let us consider $a=1$ case in (\ref{setting the Kahler parameters}). Then the partition function is
\begin{eqnarray}
\hat{\mathcal{Z}}_{M}^{A_{N-1}} =
\sum_{\text{all indices}}\prod_{s=1}^{M-1}\prod_{a=2}^{N}
\Bigl(\bar{Q}_{f,a}^{(s)}\Bigr)^{|\mu_{a}^{(s)}|}
\prod_{(i,j) \in \mu_{a}^{(s)}}\prod_{b=1}^{N}
\frac{
\theta_{1}(\tau;z_{ab}^{(s)}(i,j))
\theta_{1}(\tau;w_{ab}^{(s)}(i,j))
}{
\theta_{1}(\tau;u_{ab}^{(s)}(i,j))
\theta_{1}(\tau;v_{ab}^{(s)}(i,j))
}.
\label{partition function for some Kahler}
%(\ref{partition function for some Kahler})
\end{eqnarray}
Here we consider $b=1$ in (\ref{partition function for some Kahler}). For $b=1$ and $a>2$, (\ref{def}) becomes
\begin{eqnarray}
&&
Q_{a1}^{(s+1)} = \sqrt{\frac{t}{q}}\prod_{i=1}^{a-1}Q_{\tau_{i}}^{(s)},~
Q_{1a}^{(s)}= \sqrt{\frac{t}{q}}\prod_{i=a}^{N}Q_{\tau_{i}}^{(s)},
\nonumber \\
&&
\hat{Q}_{a1}^{(s)} =\prod_{i=1}^{a-1}Q_{\tau_{i}}^{(s)},~
\hat{Q}_{1a}^{(s)}= \prod_{i=a}^{N}Q_{\tau_{i}}^{(s)}.
\end{eqnarray}
Then, the contribution of $b=1$ in (\ref{partition function for some Kahler}) becomes $1$. Therefore, the result is as follows,
\begin{eqnarray}
\hat{\mathcal{Z}}_{M}^{A_{N-1}} =
\sum_{\text{all indices}}\prod_{s=1}^{M-1}\prod_{a=2}^{N}
\Bigl(\bar{Q}_{f,a}^{(s)}\Bigr)^{|\mu_{a}^{(s)}|}
\prod_{(i,j) \in \mu_{a}^{(s)}}\prod_{b=2}^{N}
\frac{
\theta_{1}(\tau;z_{ab}^{(s)}(i,j))
\theta_{1}(\tau;w_{ab}^{(s)}(i,j))
}{
\theta_{1}(\tau;u_{ab}^{(s)}(i,j))
\theta_{1}(\tau;v_{ab}^{(s)}(i,j))
}.
\end{eqnarray}
This partition function is the same as $\hat{\mathcal{Z}}_{M}^{A_{N-2}}$ in appropriate redefinitions of the K\"ahler parameters.
\par
Therefore, by tuning the K\"alert parameters such as (\ref{setting the Kahler parameters}), the supersymmetry of M-strings gets enhanced to $\mathcal{N}=(4,4)$.
\begin{figure}[htbp]
    \centering
    \includegraphics[width=13cm]{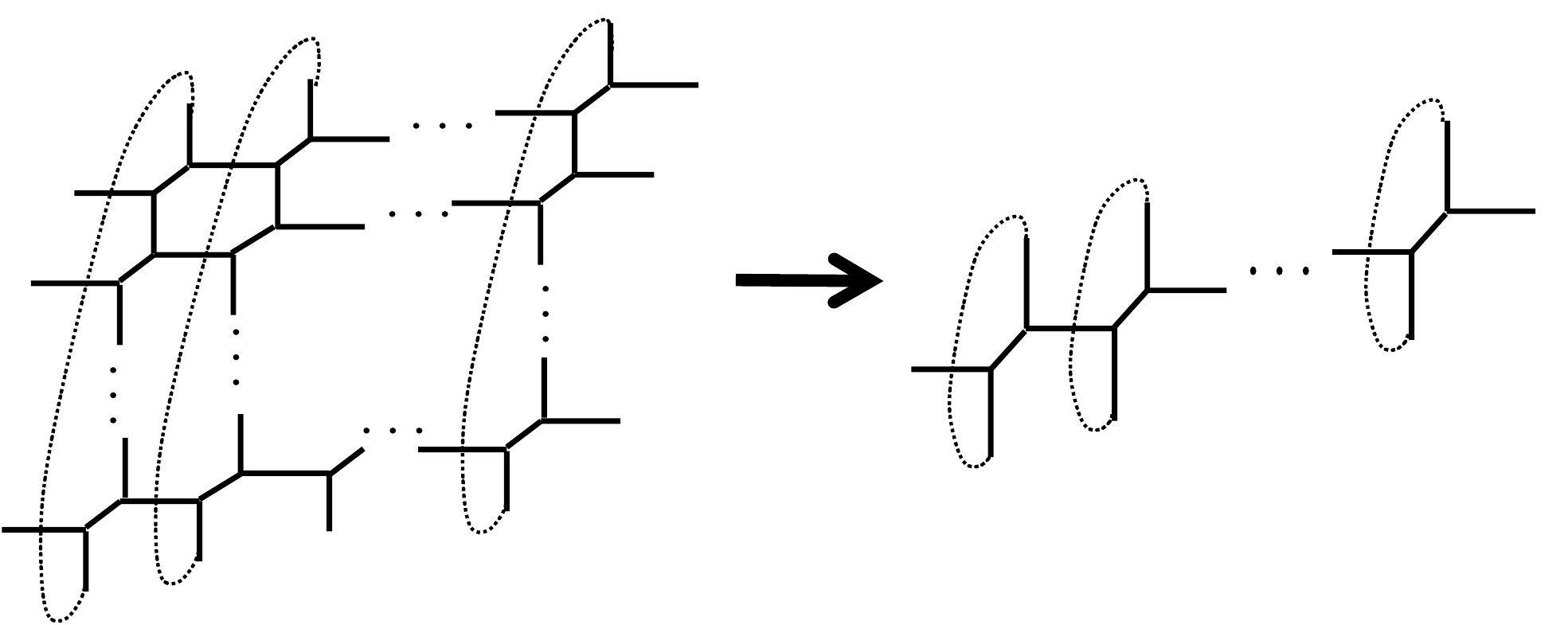}
    \caption{The enhancement of supersymmetry. By tuning the K\"ahler parameters, the web diagram reduces to the right figure trough the geometric transition.}
\end{figure}\\

\section{Discussion}
\label{sec:4}
In this paper, we have calculated the partition function of M-strings by using the refined topological vertex formalism. This partition function is consistent with \cite{Haghighat:2013tka} under the flop transition. Then we have found that, by tuning the K\"ahler parameters, the partition function of M-strings has agreed with the $\mathcal{N} = (4,4)$ partition function of M-strings. We have interpreted this result by considering the geometric transition. We have also interpreted this enhancement in terms of D1-D5 brane system on $A_{N-1}$ ALE space. 
\par
Now we want to consider another possibility of the enhancement of supersymmetry. As we mentioned in the previous subsection, the theory on $k$ D1-branes is the $\mathcal{N}=(4,0)$ $U(k)$ gauge theory with one adjoint, two chiral multiplets and two fermi multiplets. By combining $\mathcal{N}=(4,0)$ vector multiplet with adjoint multiplet, and $\mathcal{N}=(4,0)$ chiral multiplets with fermi multiplets, we may be able to obtain $\mathcal{N}=(4,4)$ vector multiplet and two chiral multiplets. Therefore,  we can expect the enhancement of supersymmetry which is different from the one in the previous section. 
\par
In order to consider, we calculate the elliptic genus which can be compared with the M-strings partition function  calculated by the refined topological string theory. The elliptic genus of the gauge theory can be calculated by using the localization method \cite{Benini:2013nda}\cite{Benini:2013xpa}\cite{Gadde:2013dda}\cite{Harvey:2014nha}\cite{Honda:2015yha}. We can write the elliptic genus of the $\mathcal{N}=(4,4)$ $U(k)$ gauge theory with two flavors,
\begin{eqnarray}
\mathcal{Z}_{Ell}(t,q,Q_{f}) &=& \oint du_{\alpha}
\Bigl(\frac{2 \pi \eta^2(\tau)}{i}\Bigr)^{k}\frac{1}{k!}
\prod_{\alpha, \beta = 1}^{k}
\frac{\theta_{1}(\tau;u_{\alpha}-u_{\beta})\theta_{1}(\tau;\epsilon_{1}+\epsilon_{2} +u_{\alpha}-u_{\beta})}{\theta_{1}(\tau;\epsilon_{1} +u_{\alpha}-u_{\beta})\theta_{1}(\tau;\epsilon_{2} +u_{\alpha}-u_{\beta})} \nonumber \\
& \times&
\prod_{\alpha=1}^{k}
\frac{ \theta_{1}(\tau ; \frac{-\epsilon_{1}+\epsilon_{2}}{2} - \xi_{f}+u_{\alpha})
\theta_{1}(\tau ; \frac{-\epsilon_{1}+\epsilon_{2}}{2} +\xi_{f} - u_{\alpha})}
{\theta_{1}(\tau;\frac{\epsilon_{1}+\epsilon_{2}}{2} -\xi_{f}+u_{\alpha})
\theta_{1}(\tau;\frac{\epsilon_{1} + \epsilon_{2}}{2} + \xi_{f} - u_{\alpha})}
\nonumber \\
& \times&
\prod_{\alpha=1}^{k}
\frac{ \theta_{1}(\tau ; \frac{-\epsilon_{1}+\epsilon_{2}}{2} +\xi_{f}+u_{\alpha})
\theta_{1}(\tau;\frac{-\epsilon_{1}+\epsilon_{2}}{2} - \xi_{f} -u_{\alpha})}
{\theta_{1}(\tau;\frac{\epsilon_{1}+\epsilon_{2}}{2}+\xi_{f}+u_{\alpha})
\theta_{1}(\tau;\frac{\epsilon_{1} + \epsilon_{2}}{2}-\xi_{f}-u_{\alpha})},
\label{enhancement}
\end{eqnarray}
where the variable $\xi_{f}$ denotes the fugacity.\par

\begin{figure}[htbp]
    \begin{center}
    \includegraphics[width=11cm]{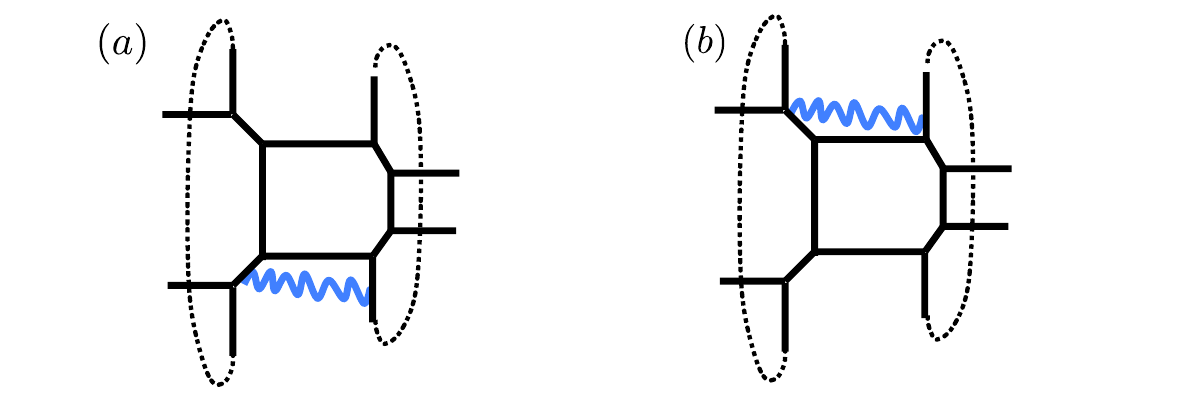}
    \caption{The contribution of strings.}
    \label{string}
  \end{center}
\end{figure}

\begin{comment}
On the other hand, we can write down the elliptic genus of the $\mathcal{N}=(4,0)$ $U(k)$ gauge theory with one adjoint and 2 flavors which should agree with the calculation result for the refined topological string theory\cite{Haghighat:2013tka},
\begin{eqnarray}
\mathcal{Z}_{Ell}(t,q,Q_{f}) &=& \oint du_{\alpha}
\Bigl(\frac{2 \pi \eta^2(\tau)}{i}\Bigr)^{k}\frac{1}{k!}
\prod_{\alpha, \beta = 1}^{k}
\frac{\theta_{1}(\tau;u_{\alpha}-u_{\beta})\theta_{1}(\tau;\epsilon_{1}+\epsilon_{2} +u_{\alpha}-u_{\beta})}{\theta_{1}(\tau;\epsilon_{1} +u_{\alpha}-u_{\beta})\theta_{1}(\tau;\epsilon_{2} +u_{\alpha}-u_{\beta})} \nonumber \\
& \times&
\prod_{\alpha=1}^{k}
\frac{ \theta_{1}(\tau ; m - \xi_{f}+u_{\alpha})
\theta_{1}(\tau ; m +\xi_{f} - u_{\alpha})}
{\theta_{1}(\tau;\frac{\epsilon_{1}+\epsilon_{2}}{2} -\xi_{f}+u_{\alpha})
\theta_{1}(\tau;\frac{\epsilon_{1} + \epsilon_{2}}{2} + \xi_{f} - u_{\alpha})}
\nonumber \\
& \times&
\prod_{\alpha=1}^{k}
\frac{ \theta_{1}(\tau ; m +\xi_{f}+u_{\alpha})
\theta_{1}(\tau; m - \xi_{f} -u_{\alpha})}
{\theta_{1}(\tau;\frac{\epsilon_{1}+\epsilon_{2}}{2}+\xi_{f}+u_{\alpha})
\theta_{1}(\tau;\frac{\epsilon_{1} + \epsilon_{2}}{2}-\xi_{f}-u_{\alpha})},
\end{eqnarray}
where the variable $m$ relates with the product of K\"ahler parameters,
\begin{eqnarray}
\frac{Q_{1}}{Q_{2}} = \frac{Q'_{2}}{Q'_{1}} = e^{2 \pi i m}
\end{eqnarray}
Thus, naively we can expect other enhancement of supersymmetry by setting the K\"ahler parameters appropriately. Indeed, 
\end{comment}
By calculating as with the reference \cite{Haghighat:2013tka}, (\ref{enhancement}) may be written as
\begin{eqnarray}
\mathcal{Z}_{Ell} &\sim&
\sum_{|\nu_{1}| + |\nu_{2}| = k} 
\prod_{
{\scriptsize
\begin{matrix}
(i_{1},j_{1})\in \nu_{1} \\
 (i_{2},j_{2})\in \nu_{1}
 \end{matrix}
 }}
\frac{\theta_{1}(\tau;(j_{1} - j_{2} + 1)\epsilon_{1} + (i_{1} - i_{2} + 1)\epsilon_{2})\theta_{1}(\tau;(j_{1}-j_{2})\epsilon_{1} +(i_{1}-i_{2})\epsilon_{2})}
{\theta_{1}(\tau;(j_{1} - j_{2})\epsilon_{1} +(i_{1} - i_{2} + 1)\epsilon_{2})\theta_{1}(\tau;(j_{1}-j_{2}+1)\epsilon_{1} + (i_{1}-i_{2})\epsilon_{2})}
 \nonumber \\ &\times&
 \prod_{(i,j)\in \nu_{1}}
 \frac
{\theta_{1}(\tau;(j-1)\epsilon_{1} +i\epsilon_{2})\theta_{1}(\tau;-j\epsilon_{1} -(i-1)\epsilon_{2})}
{\theta_{1}(\tau ; j\epsilon_{1}+i\epsilon_{2})\theta_{1}(\tau ;  (-j+1)\epsilon_{1} -(i-1)\epsilon_{2})}
 \nonumber \\ &\times&
 \prod_{(i,j)\in \nu_{1}} \frac
{\theta_{1}(\tau ; 2\xi_{f} + (j-1)\epsilon_{1}+i\epsilon_{2})\theta_{1}(\tau ; -2\xi_{f} -j\epsilon_{1} -(i-1)\epsilon_{2})}
{\theta_{1}(\tau;2\xi_{f}+j\epsilon_{1} +i\epsilon_{2})\theta_{1}(\tau;-2\xi_{f} +(-j+1)\epsilon_{1} -(i-1)\epsilon_{2})}
\nonumber \\ &\times&
\prod_{
{\scriptsize
\begin{matrix}
(i_{1},j_{1})\in \nu_{2} \\
 (i_{2},j_{2})\in \nu_{2}
 \end{matrix}
 }}
\frac{\theta_{1}(\tau;(j_{1} - j_{2} + 1)\epsilon_{1} + (i_{1} - i_{2} + 1)\epsilon_{2})\theta_{1}(\tau;(j_{1}-j_{2})\epsilon_{1} +(i_{1}-i_{2})\epsilon_{2})}
{\theta_{1}(\tau;(j_{1} - j_{2})\epsilon_{1} +(i_{1} - i_{2} + 1)\epsilon_{2})\theta_{1}(\tau;(j_{1}-j_{2}+1)\epsilon_{1} + (i_{1}-i_{2})\epsilon_{2})}
\nonumber \\ &\times&
 \prod_{(i,j)\in \nu_{2}} \frac
{\theta_{1}(\tau ; -2\xi_{f} + (j-1)\epsilon_{1}+i\epsilon_{2})\theta_{1}(\tau ; 2\xi_{f} -j\epsilon_{1} -(i-1)\epsilon_{2})}
{\theta_{1}(\tau; -2\xi_{f}+j\epsilon_{1} +i\epsilon_{2})\theta_{1}(\tau; 2\xi_{f} +(-j+1)\epsilon_{1} -(i-1)\epsilon_{2})}
\nonumber \\ &\times&
 \prod_{(i,j)\in \nu_{2}}
 \frac
{\theta_{1}(\tau;(j-1)\epsilon_{1} +i\epsilon_{2})\theta_{1}(\tau;-j\epsilon_{1} -(i-1)\epsilon_{2})}
{\theta_{1}(\tau ; j\epsilon_{1}+i\epsilon_{2})\theta_{1}(\tau ;  (-j+1)\epsilon_{1} -(i-1)\epsilon_{2})}
.
\label{locus}
\end{eqnarray}
On the other hand, we rewrite the partition function calculated by the refined topological string,
\begin{eqnarray}
\hat{\mathcal{Z}} &=& \sum_{\nu_{1},\nu_{2}} 
\Bigl({Q}_{1}{Q'}_{1}{Q}_{2}{Q'}_{2}Q_{b}\Bigr)^{\frac{|\nu_{1}|+|\nu_{2}|}{2}}
\nonumber \\ &&\times
 \prod_{(i,j) \in \nu_{1}}
 \frac{
 \theta_{1}(\tau ; u^{1}_{ij})\theta_{1}(\tau ; {U}^{2f}_{ij})\theta_{1}(\tau ; u'^{1}_{ij})\theta_{1}(\tau ; {U'}^{2f}_{ij})
 }{
 \theta_{1}(\tau ; \hat{y}^{f}_{ij})\theta_{1}(\tau ; v^{\nu_{1}}_{ij})\theta_{1}(\tau ; Y^{f}_{ij})\theta_{1}(\tau ; V^{\nu_{1}}_{ij})
 } 
 \nonumber \\ &&\times
\prod_{(i,j) \in \nu_{2}}\frac{\theta_{1}(\tau ; \tilde{u}'^{2}_{ij})\theta_{1}(\tau ; \tilde{U}'^{1f}_{ij})\theta_{1}(\tau ; \tilde{u}^{2}_{ij})\theta_{1}(\tau ; \tilde{U}^{1f}_{ij})}{\theta_{1}(\tau ; w^{f}_{ij})\theta_{1}(\tau ; v^{\nu_{2}}_{ij})\theta_{1}(\tau ; \hat{W}^{f}_{ij})\theta_{1}(\tau ; V^{\nu_{2}}_{ij})}
,
\end{eqnarray}
where we define
\begin{eqnarray}
\mathrm{e}^{2 \pi i \hat{y}^{I}_{ij}} &=& Q_{I}^{-1}t^{-\nu_{1,i}+j-1}q^{-\nu^{t}_{2,j}+i},~~
\mathrm{e}^{2 \pi i \hat{W}^{I}_{ij}} ~~=~~ Q_{I}^{-1}t^{\nu_{2,i}-j}q^{\nu^{t}_{1,j}-i+1}.
\end{eqnarray}
Then, we define the partition function of $k$ M-strings as follows,
\begin{eqnarray}
\hat{\mathcal{Z}}^{k}_{M-strings}
&=&
\sum_{|\nu_{1}| + |\nu_{2}| =k}
 \prod_{(i,j) \in \nu_{1}}
 \frac{
 \theta_{1}(\tau ; u^{1}_{ij})\theta_{1}(\tau ; {U}^{2f}_{ij})\theta_{1}(\tau ; u'^{1}_{ij})\theta_{1}(\tau ; {U'}^{2f}_{ij})
 }{
 \theta_{1}(\tau ; \hat{y}^{f}_{ij})\theta_{1}(\tau ; v^{\nu_{1}}_{ij})\theta_{1}(\tau ; Y^{f}_{ij})\theta_{1}(\tau ; V^{\nu_{1}}_{ij})
 } 
 \nonumber \\ &&\times
\prod_{(i,j) \in \nu_{2}}\frac{\theta_{1}(\tau ; \tilde{u}'^{2}_{ij})\theta_{1}(\tau ; \tilde{U}'^{1f}_{ij})\theta_{1}(\tau ; \tilde{u}^{2}_{ij})\theta_{1}(\tau ; \tilde{U}^{1f}_{ij})}{\theta_{1}(\tau ; w^{f}_{ij})\theta_{1}(\tau ; v^{\nu_{2}}_{ij})\theta_{1}(\tau ; \hat{W}^{f}_{ij})\theta_{1}(\tau ; V^{\nu_{2}}_{ij})}
\label{rewrite1}
\end{eqnarray}
Then, we find that (\ref{locus}) is similar to  (\ref{rewrite1}). Indeed, for $\nu_{1}=\emptyset$ or $\nu_{2}=\emptyset$, (\ref{locus}) agrees with (\ref{rewrite1}) by tuning the K\"ahler parameters  as follows,
\begin{eqnarray}
Q_{1}=Q_{2}=\sqrt{\frac{1}{tq}},~Q'_{1}=Q'_{2}=\sqrt{tq}.
\label{mass}
\end{eqnarray}
(\ref{mass}) means that we set the mass parameter $m$ to zero. The partition function  (\ref{rewrite1}) in $\nu_{1}=\emptyset$ or $\nu_{2}=\emptyset$ means that we take the contribution of the strings which are stretched along the lower framing (Fig. \ref{string} (a)) or the upper framing (Fig. \ref{string} (b)).
\par
Thus, we can expect the enhancement of supersymmetry which is different from the one in the previous section.
\par
However, we could not obtain the elliptic genus which agreed with the partition function of M-strings perfectly.  This result about the elliptic genus depends on taking some poles. Therefore, we would like to consider this problem as a future work.

\begin{comment}
We now interpret this result. In general, the enhancement of supersymmetry does not occur because of $A_{1}$ ALE space and the mass deformations. However, when we set the K\"ahler parameters,  we can set the physical mass parameter to zero. Thus,  the supersymmetry breaking due to the mass parameter is resolved. Moreover, according to the reference \cite{Okuyama:2005gq}, the quiver gauge theory which corresponds to M-strings is $\mathcal{N}=(4,0)$ $U(k)$ gauge theory with one adjoint and two fundamental hypermultipelts. Thus if we can construct the $\mathcal{N}=(4,4)$ vector multiplet by combining the $\mathcal{N}=(4,0)$ vector multiplet and adjoint multiplet, we can obtain $\mathcal{N}=(4,4)$ theory with fundamental multiplets. Thus, we can expect that the $\mathcal{N}=(4,0)$ supersymmetry gets enhanced to the $\mathcal{N}=(4,4)$ supersymmetry by setting the K\"ahler parameters.\par
\end{comment}

\section*{Acknowledgement}
I would like to thank Satoshi Yamaguchi for several discussion and check the manuscript. I would also like to thank to Hironori Mori for some comments and discussion.

\newpage

\renewcommand{\thesection}{A}
\label{appA}
\section{Definition and Formula}
The theta function, the refined topological vertex, and its gluing factor are defined as follows:
\begin{itemize}
\item The theta function and its property
\begin{eqnarray}
\theta_{1}(\tau ; z) &=& -i e^{\frac{i \pi \tau}{4}}e^{i \pi z} \prod_{n=0}^{\infty} \Bigl\{(1-e^{2 \pi i(n+1) \tau})(1-e^{2 \pi i(n+1) \tau}e^{2 \pi i z})(1-e^{2 \pi i n \tau}e^{-2 \pi i z}) \Bigr\}  \nonumber \\
&=& -i e^{\frac{i \pi \tau}{4}}(e^{i \pi z} - e^{-i \pi z}) 
\nonumber \\ &&\times
 \prod_{n=0}^{\infty} \Bigl\{(1-e^{2 \pi i(n+1) \tau})(1-e^{2 \pi i(n+1) \tau}e^{2 \pi i z})(1-e^{2 \pi i(n+1) \tau}e^{-2 \pi i z}) \Bigr\}
 \\
 \theta_{1}(\tau ; -z) &=& -\theta_{1}(\tau ; z)
\end{eqnarray}
\item The refined topological vertex
\begin{eqnarray}
C_{\lambda \mu \nu}(t,q) &=& t^{-\frac{||\mu^{t}||^{2}}{2}}q^{\frac{||\mu||^2 + ||\nu||^{2}}{2}} \tilde{Z}_{\nu}(t,q)\nonumber \\
&& \times \sum_{\eta}\Bigl(\frac{q}{t}\Bigr)^{\frac{|\eta| + |\lambda| - |\mu|}{2}}  s_{\lambda^{t}/\eta}(t^{-\rho}q^{-\nu})s_{\mu/\eta}(t^{-\nu^{t}}q^{-\rho}) \\
\tilde{Z}_{\nu}(t,q) &=& \prod_{(i,j) \in \nu}(1-q^{\nu_{i}-j}t^{\nu_{j}^{t} -i +1})^{-1} 
\nonumber
\end{eqnarray}
\item The gluing factors
\begin{eqnarray}
f_{\mu} (t,q) = (-1)^{|\mu|}q^{-\frac{||\mu||^2}{2}}t^{\frac{||\mu^{t}||^2}{2}},~~
\tilde{f}_{\mu} (t,q) = (-1)^{|\mu|}\Bigl(\frac{t}{q}\Bigr)^{\frac{|\mu|}{2}}q^{-\frac{||\mu||^2}{2}}t^{\frac{||\mu^{t}||^2}{2}},
\end{eqnarray}
\end{itemize}
We can use some formulas to calculate the partition function and the recursion formula that we will explain in the next subsection:
\begin{itemize} 
\item Some formulas about schur polynomial
\begin{eqnarray}
s_{\lambda/\mu}(\alpha \bold{x}) &=& \alpha^{|\lambda|-|\mu|}s_{\lambda/\mu}(\bold{x}) \\
\sum_{\eta}s_{\eta/\lambda}(\bold{x})s_{\eta/\mu}(\bold{y}) &=& \prod_{i,j=1}^{\infty}(1-x_{i}y_{j})^{-1}\sum_{\tau}s_{\mu/\tau}(\bold{x})s_{\lambda/\tau}(\bold{y}) \\
\sum_{\eta}s_{\eta^{t}/\lambda}(\bold{x})s_{\eta/\mu}(\bold{y}) &=& \prod_{i,j=1}^{\infty}(1+x_{i}y_{j})\sum_{\tau}s_{\mu^{t}/\tau}(\bold{x})s_{\lambda^{t}/\tau^{t}}(\bold{y})
\end{eqnarray}
\item Normalization
\begin{eqnarray}
\prod_{i,j=1}^{\infty} \frac{1-Qq^{\nu_{i}-j}t^{\mu_{j}^{t}-i+1}}{1-Qq^{-j}t^{-i+1}} &=& \prod_{(i,j) \in \nu}(1-Qq^{\nu_{i}-j}t^{\mu_{j}^{t}-i+1})\prod_{(i,j) \in \mu}(1-Qq^{-\mu_{i}+j-1}t^{-\nu_{j}^{t}+i}) ~~~~~~~~~~~~\\
\prod_{i,j=1}^{\infty} \frac{1-Qt^{\nu_{j}^{t}-i+\frac{1}{2}}q^{-j+\frac{1}{2}}}{1-Qt^{-i+\frac{1}{2}}q^{-j+\frac{1}{2}}} &=& \prod_{(i,j) \in \nu}(1-Qq^{-j+\frac{1}{2}}t^{i-\frac{1}{2}}) \\
\prod_{i,j=1}^{\infty} \frac{1-Qq^{\nu_{i}-j+\frac{1}{2}}t^{-i+\frac{1}{2}}}{1-Qq^{-j+\frac{1}{2}}t^{-i+\frac{1}{2}}} &=&\prod_{(i,j) \in \nu}(1-Qq^{j-\frac{1}{2}}t^{-i+\frac{1}{2}})
\end{eqnarray}
%\item The identity
%\begin{eqnarray}
%\prod_{n=0}^{\infty}(1-Ax^{n}) = \prod_{n=0}^{\infty}(1-Ax^{-n-1})^{-1}
%\end{eqnarray}
\item Some factor calculation
\begin{eqnarray}
&&\sum_{(i,j)\in \nu} (\nu_{i} - j + \nu_{j}^{t} - i +1) = \frac{||\nu||^{2}}{2} + \frac{||\nu^{t}||^{2}}{2},~~\sum_{(i,j) \in \nu}(j-i) = \frac{||\nu||^{2}}{2}  \\
&&\sum_{(i,j)\in \nu}\mu^{t}_{j} = \sum_{(i,j) \in \mu}\nu_{j}^{t},~~\sum_{(i,j) \in \nu}\nu^{t}_{j} = ||\nu^{t}||^{2}
 \\
&&\sum_{(i,j) \in \nu}(\nu^{t}_{j}-i) = \frac{||\nu^{t}||^{2}}{2}-\frac{{|\nu|}}{2},~~\sum_{(i,j) \in \nu}(\nu_{i}-j) = \frac{||\nu||^{2}}{2}-\frac{{|\nu|}}{2}
\\
&&\sum_{(i,j) \in \mu}i = \frac{||\mu^{t}||}{2} + \frac{|\mu|}{2},~~
\sum_{(i,j) \in \mu}j = \frac{||\mu||}{2} + \frac{|\mu|}{2}
\end{eqnarray}
\end{itemize}

\renewcommand{\thesection}{B}
\section{Recursion Formula and Flop Transition}
In this section, we calculate the building block about two cases, and  we derive the recursion formula. Then, we discuss how to connect two building blocks under the flop transition.

\subsection{Simple case}
we consider the following web diagrams, (a) and (b).\\
\begin{figure}[htbp]
\renewcommand{\thefigure}{B.1}
    \begin{center}
    \includegraphics[clip,width=7cm]{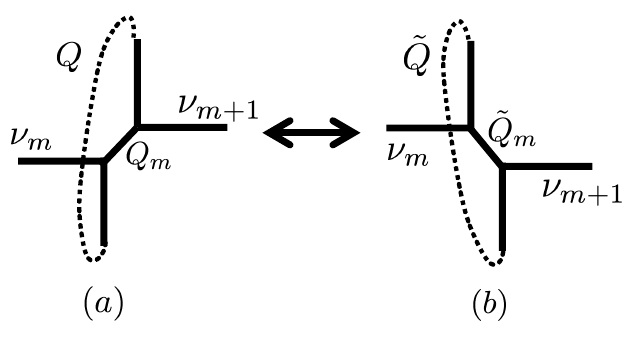}
        \label{building2}
    \caption{The simple web diagrams where $Q_{m}$ and $QQ_{m}$ denote the mass deformation and the radius of compactification, respectively.  The preferred direction is along the horizontal.}
  \end{center}
\end{figure}\\
Then, we can write the partition function for the web diagram (a),
\begin{eqnarray}
\mathcal{Z}^{(a)}_{\nu_{m}\nu_{m+1}} &=& \sum_{\mu_{m},\mu}(-Q_{m})^{|\mu_{m}|}(-Q)^{|\mu|}
C_{\mu \mu_{m} \nu_{m}^{t}}(t,q)C_{\mu^{t} \mu_{m}^{t} \nu_{m+1}}(q,t)
\nonumber \\
&=& q^{\frac{1}{2}||\nu_{m}^{t}||}t^{\frac{1}{2}||\nu_{m+1}||}\tilde{Z}_{\nu_{m}^{t}}(t,q)\tilde{Z}_{\nu_{m+1}}(q,t)
\nonumber \\ &&\times
\sum_{\mu_{m},\mu,\eta_{1,2}}(-Q_{m})^{|\mu_{m}|}(-Q)^{|\mu|}
s_{\mu/\eta_{1}}(t^{-\rho+\frac{1}{2}}q^{-\nu_{m}^{t}-\frac{1}{2}})s_{\mu^{t}/\eta_{2}}(q^{-\rho+\frac{1}{2}}t^{-\nu_{m+1}-\frac{1}{2}})
\nonumber \\ &&
~~~~~~~~~~~~~~~~~~~~~~~~~~~~~~~~\times
s_{\mu_{m}^{t}/\eta_{1}}(q^{-\rho}t^{-\nu_{m}})s_{\mu_{m}/\eta_{2}}(t^{-\rho}q^{-\nu_{m+1}^{t}}).
\end{eqnarray}
In order to calculate this, we consider the following equation,
\begin{eqnarray}
A(x_{1},x_{2},x_{3},x_{4}) = \sum_{\mu_{m},\mu,\eta_{1,2}}
\alpha_{m}^{|\mu_{m}|}\alpha^{|\mu|}
s_{\mu/\eta_{1}}(x_{1})s_{\mu^{t}/\eta_{2}}(x_{2})
s_{\mu_{m}^{t}/\eta_{1}}(x_{3})s_{\mu_{m}/\eta_{2}}(x_{4}).
\end{eqnarray}
We can use the formula in the reference \cite{Haghighat:2013gba}.
Then we obtain
\begin{eqnarray}
\hat{\mathcal{Z}}^{(a)}_{\nu_{m}\nu_{m+1}}&:=& 
\mathcal{Z}^{(a)}_{\nu_{m}\nu_{m+1}}/\mathcal{Z}^{(a)}_{\emptyset\emptyset}
\nonumber \\
&=&
q^{\frac{1}{2}||\nu_{m}^{t}||}t^{\frac{1}{2}||\nu_{m+1}||}
\nonumber \\ &&\times
\prod_{n=0}^{\infty}
\prod_{(i,j)\in \nu_{m}}
\frac{
(1-Q_{m}Q_{\tau}^{n}t^{-\nu_{m,i}+j-\frac{1}{2}}q^{-\nu_{m+1,j}^{t}+i-\frac{1}{2}})
(1-Q_{m}^{-1}Q_{\tau}^{n+1}t^{\nu_{m,i}-j+\frac{1}{2}}q^{\nu_{m+1,j}^{t}-i+\frac{1}{2}})
}{
(1-Q_{\tau}^{n}t^{\nu_{m,i}-j+1}q^{\nu^{t}_{m,j}-i})(1-Q_{\tau}^{n+1}t^{-\nu_{m,i}+j}q^{-\nu^{t}_{m,j}+i-1})} 
\nonumber \\ &&\times
\prod_{(i,j)\in \nu_{m+1}}
\frac{
(1-Q_{m}Q_{\tau}^{n}t^{\nu_{m+1,i}-j+\frac{1}{2}}q^{\nu_{m,j}^{t}-i+\frac{1}{2}})
(1-Q_{m}^{-1}Q_{\tau}^{n+1}t^{-\nu_{m+1,i}+j-\frac{1}{2}}q^{-\nu_{m,j}^{t}+i-\frac{1}{2}})
}{
(1-Q_{\tau}^{n}t^{\nu_{m+1,i}-j}q^{\nu^{t}_{m+1,j}-i+1})(1-Q_{\tau}^{n+1}t^{-\nu_{m+1,i}+j-1}q^{-\nu^{t}_{m,j}+i})},
\nonumber \\
\label{build1norm}
\end{eqnarray}
where we define $Q_{\tau}=Q_{m}Q$.\par
Next we calculate the partition function for the web diagram (b). This calculation is the same as the partition function $\mathcal{Z}^{(a)}_{\nu_{m}\nu_{m+1}}$. Then we obtain
\begin{eqnarray}
\mathcal{Z}^{(b)}_{\nu_{m}\nu_{m+1}} &=&
 \sum_{\mu_{m},\mu}(-\tilde{Q}_{m})^{|\mu_{m}|}(-\tilde{Q})^{|\mu|}
C_{\mu_{m} \mu \nu_{m}^{t}}(t,q)C_{\mu^{t}_{m} \mu^{t} \nu_{m+1}}(q,t)
\nonumber \\
&=& q^{\frac{1}{2}||\nu_{m}^{t}||}t^{\frac{1}{2}||\nu_{m+1}||}\tilde{Z}_{\nu_{m}^{t}}(t,q)\tilde{Z}_{\nu_{m+1}}(q,t)
\nonumber \\ &&\times
\sum_{\mu_{m},\mu,\eta_{1,2}}(-\tilde{Q}_{m})^{|\mu_{m}|}(-\tilde{Q})^{|\mu|}
s_{\mu_{m}/\eta_{1}}(t^{-\rho+\frac{1}{2}}q^{-\nu_{m}^{t}-\frac{1}{2}})
s_{\mu_{m}^{t}/\eta_{2}}(q^{-\rho+\frac{1}{2}}t^{-\nu_{m+1}-\frac{1}{2}})
\nonumber \\ &&
~~~~~~~~~~~~~~~~~~~~~~~~~~~~~~~~\times
s_{\mu^{t}/\eta_{1}}(q^{-\rho}t^{-\nu_{m}})s_{\mu/\eta_{2}}(t^{-\rho}q^{-\nu_{m+1}^{t}}),
\end{eqnarray}
and
\begin{eqnarray}
\hat{\mathcal{Z}}^{(b)}_{\nu_{m}\nu_{m+1}}&:=& 
\mathcal{Z}^{(b)}_{\nu_{m}\nu_{m+1}}/\mathcal{Z}^{(b)}_{\emptyset\emptyset}
\nonumber \\
&=&
q^{\frac{1}{2}||\nu_{m}^{t}||}t^{\frac{1}{2}||\nu_{m+1}||}
\nonumber \\ &&\times
\prod_{n=0}^{\infty}
\prod_{(i,j)\in \nu_{m}}
\frac{
(1-\tilde{Q}_{m}\tilde{Q}_{\tau}^{n}t^{\nu_{m,i}-j+\frac{1}{2}}q^{\nu_{m+1,j}^{t}-i+\frac{1}{2}})
(1-\tilde{Q}_{m}^{-1}\tilde{Q}_{\tau}^{n+1}t^{-\nu_{m,i}+j-\frac{1}{2}}q^{-\nu_{m+1,j}^{t}+i-\frac{1}{2}})
}{
(1-\tilde{Q}_{\tau}^{n}t^{\nu_{m,i}-j+1}q^{\nu^{t}_{m,j}-i})(1-\tilde{Q}_{\tau}^{n+1}t^{-\nu_{m,i}+j}q^{-\nu^{t}_{m,j}+i-1})} 
\nonumber \\ &&\times
\prod_{(i,j)\in \nu_{m+1}}
\frac{
(1-\tilde{Q}_{m}\tilde{Q}_{\tau}^{n}t^{-\nu_{m+1,i}+j-\frac{1}{2}}q^{-\nu_{m,j}^{t}+i-\frac{1}{2}})
(1-\tilde{Q}_{m}^{-1}\tilde{Q}_{\tau}^{n+1}t^{\nu_{m+1,i}-j+\frac{1}{2}}q^{\nu_{m,j}^{t}-i+\frac{1}{2}})
}{
(1-\tilde{Q}_{\tau}^{n}t^{\nu_{m+1,i}-j}q^{\nu^{t}_{m+1,j}-i+1})
(1-\tilde{Q}_{\tau}^{n+1}t^{-\nu_{m+1,i}+j-1}q^{-\nu^{t}_{m,j}+i})}.
\nonumber \\
\end{eqnarray}
These results agree with each other under the flop transition (see Fig. \ref{building2}). The K\"ahler parameters are related as follows,
\begin{eqnarray}
Q_{m} = \tilde{Q}^{-1}_{m},~~Q_{\tau}=\tilde{Q}_{\tau}(Q=\tilde{Q}_{m}\tilde{Q}\tilde{Q}_{m}).
\end{eqnarray}

\subsection{Little complicated case}
\begin{figure}[htbp]
\renewcommand{\thefigure}{B.2}
\center
    \includegraphics[clip,width=7cm]{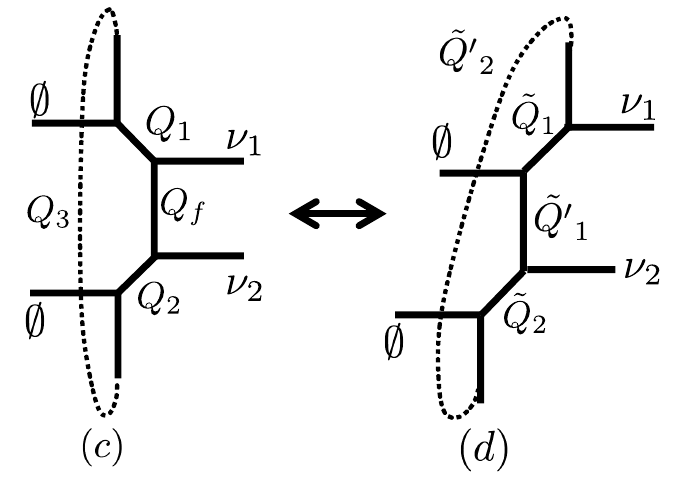}
        \label{flop transition}
    \caption{The flop transition for the little complicated web diagram.}
\end{figure}

In this subsection we calculate the following web diagram (see Fig. {\ref{flop transition}}).
\par
We have already written the partition function for the web diagram (c) in section 3. In order to calculate this, we consider the following equation,
\begin{eqnarray}
F(x_{1},x_{2},...,x_{8}) &=& \sum_{\mu_{f,1,2,3},~\eta_{1,2,3,4}}  \alpha_{f}^{|\mu_{f}|}\alpha_{1}^{|\mu_{1}|}\alpha_{2}^{|\mu_{2}|}\alpha_{3}^{|\mu_{3}|}\beta_{1}^{|\eta_{1}|}\beta_{2}^{|\eta_{2}|}\beta_{3}^{|\eta_{3}|}\beta_{4}^{|\eta_{4}|} 
\nonumber  \\ &&\times
s_{\mu_{1}/\eta_{1}}(\bold{x}_{1})s_{\mu_{3}/\eta_{1}}(\bold{x}_{2})
s_{\mu_{1}^{t}/\eta_{2}}(\bold{x}_{3})s_{\mu_{f}^{t}/\eta_{2}}(\bold{x}_{4})
\nonumber  \\ &&\times
s_{\mu_{f}^{t}/\eta_{3}}(\bold{x}_{5})s_{\mu_{2}^{t}/\eta_{3}}(\bold{x}_{6})
s_{\mu_{3}/\eta_{4}}(\bold{x}_{7})s_{\mu_{2}/\eta_{4}}(\bold{x}_{8})
.
\label{definition of F}
\end{eqnarray}
We calculate (\ref{definition of F}) by using some formulas in Appendix A. Then we obtain
\par
\begin{eqnarray}
F(\bold{x}_{i}) &=& \prod_{i,j=1}^{\infty}
\frac{
(1+\alpha_{1}x_{1}^{i}x_{3}^{j})(1+\alpha_{2}x_{6}^{i}x_{8}^{j})
(1+\alpha_{1}^{-1}\Lambda \beta_{3}\beta_{4}x_{2}^{i}x_{4}^{j})(1+\alpha_{2}^{-1}\Lambda\beta_{1}\beta_{2}x_{5}^{i}x_{7}^{j})
}{
(1-\alpha_{3}x_{2}^{i}x_{7}^{j})(1-\alpha_{f}x_{4}^{i}x_{5}^{j})
(1-\alpha_{f}^{-1}\Lambda\beta_{1}\beta_{4}x_{3}^{i}x_{6}^{j})(1-\alpha_{3}^{-1}\Lambda \beta_{2}\beta_{3}x_{1}^{i}x_{8}^{j})
}
\nonumber \\ &&\times
\frac{
(1+\alpha_{1}\alpha_{3}\beta_{1}x_{3}^{i}x_{7}^{j})(1+\alpha_{2}\alpha_{3}\beta_{4}x_{2}^{i}x_{6}^{j})
(1+\alpha_{1}^{-1}\alpha_{2}^{-1}\Lambda\beta_{2}x_{4}^{i}x_{8}^{j})(1+\alpha_{2}^{-1}\alpha_{3}^{-1}\Lambda\beta_{2}x_{1}^{i}x_{5}^{j})
}{
(1-\Lambda\beta_{2}\beta_{3}\beta_{4}x_{1}^{i}x_{2}^{j})(1-\Lambda\beta_{1}\beta_{3}\beta_{4}x_{3}^{i}x_{4}^{j})
(1-\Lambda\beta_{1}\beta_{2}\beta_{4}x_{5}^{i}x_{6}^{j})(1-\Lambda\beta_{1}\beta_{2}\beta_{3}x_{7}^{i}x_{8}^{j})
} 
\nonumber \\ &&\times
\frac{
(1+\alpha_{1}\Lambda x_{1}^{i}x_{3}^{j})(1+\alpha_{2}\Lambda x_{6}^{i}x_{8}^{j})
(1+\alpha_{1}^{-1}\Lambda^{2} \beta_{3}\beta_{4}x_{2}^{i}x_{4}^{j})(1+\alpha_{2}^{-1}\Lambda^{2}\beta_{1}\beta_{2}x_{5}^{i}x_{7}^{j})
}{
(1-\alpha_{3}\Lambda x_{2}^{i}x_{7}^{j})(1-\alpha_{f}\Lambda x_{4}^{i}x_{5}^{j})
(1-\alpha_{f}^{-1}\Lambda^{2}\beta_{1}\beta_{4}x_{3}^{i}x_{6}^{j})(1-\alpha_{3}^{-1}\Lambda^{2} \beta_{2}\beta_{3}x_{1}^{i}x_{8}^{j})
}
\nonumber \\ &&\times
\frac{
(1+\alpha_{1}\alpha_{3}\Lambda \beta_{1}x_{3}^{i}x_{7}^{j})(1+\alpha_{2}\alpha_{3}\beta_{4}\Lambda x_{2}^{i}x_{6}^{j})
(1+\alpha_{1}^{-1}\alpha_{2}^{-1}\Lambda^{2}\beta_{2}x_{4}^{i}x_{8}^{j})(1+\alpha_{2}^{-1}\alpha_{3}^{-1}\Lambda^{2}\beta_{2}x_{1}^{i}x_{5}^{j})
}{
(1-\Lambda^{2}\beta_{2}\beta_{3}\beta_{4}x_{1}^{i}x_{2}^{j})(1-\Lambda^{2}\beta_{1}\beta_{3}\beta_{4}x_{3}^{i}x_{4}^{j})
(1-\Lambda^{2}\beta_{1}\beta_{2}\beta_{4}x_{5}^{i}x_{6}^{j})(1-\Lambda^{2}\beta_{1}\beta_{2}\beta_{3}x_{7}^{i}x_{8}^{j})
}  \nonumber \\ &&\times 
F(\Lambda \bold{x}_{i})
\nonumber \\ 
&=:& G(\bold{x})F(\Lambda \bold{x}_{i}).\nonumber \\
&&(\Lambda := \alpha_{1} \alpha_{2} \alpha_{3} \alpha_{f}) ~~~~~~~~~~~~~~~~~~~~~~~~~~~~~~~~~~~~~~~~~~~~~~~~~~~~~~~~~~~~~~~~~~~~~~~~~~~~~~~~
\end{eqnarray}
Thus, we get
\begin{eqnarray}
F(x_{i})=\prod_{i=1}^{\infty}G(\Lambda^{i-1}\bold{x})\lim_{n \to \infty}F(\Lambda^{n}\bold{x}_{i}).
\label{recursion}
\end{eqnarray}
In order to consider the non-zero value for $\lim_{n \to \infty}F(\Lambda^{n}\bold{x}_{i})$, we set the representation of some schur functions to trivial. In order to do this,  we restrict (\ref{definition of F}) to
\begin{eqnarray}
\mu_{3}^{t} = \eta_{1},~~\mu_{1}^{t} =\eta_{1},~~\mu_{f} = \eta_{2},~~\mu_{1} = \eta_{2}, \nonumber \\
\mu_{2} = \eta_{3},~~\mu_{f} = \eta_{3},~~\mu_{2}^{t} = \eta_{4},~~\mu_{3}^{t} = \eta_{4}.
\end{eqnarray}
Then we obtain\footnote{We assume $\lim_{n \to \infty} \Lambda^{n} = 0$.}
\begin{eqnarray}
\lim_{n \to \infty}F(\Lambda^{n}\bold{x}_{i}) &=& \sum_{\eta}\Lambda^{|\eta|}\Gamma^{|\eta|}\nonumber \\
&=& \prod_{k=1}^{\infty}(1-\Lambda^k \Gamma^k)^{-1}. \\
&&(\Gamma := \beta_{1}\beta_{2}\beta_{3}\beta_{4}) \nonumber
\end{eqnarray}
Therefore, we conclude
\begin{eqnarray}
F(\bold{x}_{i}) = \prod_{k=1}^{\infty}(1-\Lambda^k \Gamma^k)^{-1} \prod_{i=1}^{\infty}G(\Lambda^{i-1}\bold{x}).
\end{eqnarray}
\par
Next we calculate the partition function for the web diagram (d). This partition function has been calculated in the reference \cite{Haghighat:2013tka}. Then, in the above definition of the K\"ahler parameters, we find
\begin{eqnarray}
\hat{\mathcal{Z}}_{\nu_{1}\nu_{2}}
&=& 
\tilde{Z}_{\nu_{1}}(q,t)\tilde{Z}_{\nu_{2}}(q,t)t^{\frac{1}{2}(||\nu_{1}||^2+||\nu_{2}||^2)} 
 \nonumber \\ &&\times
\prod_{n=0}^{\infty}
\prod_{(i,j) \in \nu_{1}} 
\Biggl \{
\frac{
(1-\tilde{Q}_{1}\Lambda^{n}t^{\nu_{1,i}-j+\frac{1}{2}}q^{-i+\frac{1}{2}})
(1-\tilde{Q}_{1}^{-1}\Lambda^{n+1}t^{-\nu_{1,i}+j-\frac{1}{2}}q^{i-\frac{1}{2}})
}{
(1-\Lambda^{n+1}t^{-\nu_{1,i}+j-1}q^{-\nu^{t}_{1,j}+i})
(1-\Lambda^{n+1}t^{\nu_{1,i}-j}q^{\nu^{t}_{1,j}-i+1})
}
 \Biggr \}
\nonumber \\ &&\times
\frac{
(1-\tilde{Q}_{1}\tilde{Q}_{2}\tilde{Q'}_{1}\Lambda^{n}t^{\nu_{1,i}-j+\frac{1}{2}}q^{-i+\frac{1}{2}})
(1-\tilde{Q}_{1}^{-1}\tilde{Q}_{2}^{-1}\tilde{Q'}_{1}^{-1}\Lambda^{n+1}t^{-\nu_{1,i}+j-\frac{1}{2}}q^{i-\frac{1}{2}})
}{
(1-\tilde{Q}_{1}\tilde{Q'}_{1}\Lambda^{n}t^{\nu_{1,i}-j}q^{\nu^{t}_{2,j}-i+1})
(1-\tilde{Q}_{2}\tilde{Q'}_{2}\Lambda^{n}t^{-\nu_{1,i}+j-1}q^{-\nu_{2,j}^{t} +i})
}
\nonumber \\ &&\times
\prod_{(i,j) \in \nu_{2}} 
\Biggl \{
\frac{
(1-\tilde{Q}_{2}\Lambda^{n}t^{\nu_{2,i}-j+\frac{1}{2}}q^{-i+\frac{1}{2}})
(1-\tilde{Q}_{2}^{-1}\Lambda^{n+1}t^{-\nu_{2,i}+j-\frac{1}{2}}q^{i-\frac{1}{2}})
}{
(1-\Lambda^{n+1}t^{-\nu_{2,i}+j-1}q^{-\nu_{2,j}^{t}+i})
(1-\Lambda^{n+1}t^{\nu_{2,i}-j}q^{\nu_{2,j}^{t}-i+1})
} 
\nonumber \\ &&\times
\frac{
(1-\tilde{Q}_{1}\tilde{Q}_{2}\tilde{Q'}_{2}\Lambda^{n}t^{\nu_{2,i}-j+\frac{1}{2}}q^{-i+\frac{1}{2}})
(1-\tilde{Q}_{1}^{-1}\tilde{Q}_{2}^{-1}\tilde{Q'}_{2}^{-1}\Lambda^{n+1}t^{-\nu_{2,i}+j-\frac{1}{2}}q^{i-\frac{1}{2}})
}{
(1-\tilde{Q}_{1}\tilde{Q'}_{1}\Lambda^{n}t^{-\nu_{2,i}+j-1}q^{-\nu^{t}_{1,j}+i})
(1-\tilde{Q}_{2}\tilde{Q'}_{2}\Lambda^{n}t^{\nu_{2,i}-j}q^{\nu^{t}_{1,j}-i+1})
}
 \Biggr \}.
 \nonumber \\
 \end{eqnarray}
Again these results agree with each other under the flop transition. The K\"ahler parameters are related as follows,
\begin{eqnarray}
Q_{1}=\tilde{Q}_{1}^{-1},~~Q_{f} =  \tilde{Q}_{1}\tilde{Q'}_{1},~~
Q_{2} = \tilde{Q}_{2},~~Q_{3} = \tilde{Q}_{1}\tilde{Q'}_{2}.
\end{eqnarray}


\newpage



\begin{thebibliography}{99}

\bibitem{Haghighat:2013gba}
  B.~Haghighat, A.~Iqbal, C.~Kozcaz, G.~Lockhart and C.~Vafa,
  ``M-Strings,''
  Commun.\ Math.\ Phys.\  {\bf 334} (2015) 2,  779
  [arXiv:1305.6322 [hep-th]].
  %%CITATION = ARXIV:1305.6322;%%
  %26 citations counted in INSPIRE as of 21 May 2015

\bibitem{Hohenegger:2013ala} 
  S.~Hohenegger and A.~Iqbal,
  ``M-strings, elliptic genera and $\mathcal{N} = 4$ string amplitudes,''
  Fortsch.\ Phys.\  {\bf 62}, 155 (2014)
  [arXiv:1310.1325 [hep-th]].
  %%CITATION = ARXIV:1310.1325;%%
  %10 citations counted in INSPIRE as of 19 May 2015
  
  %\cite{Haghighat:2013tka}
  \bibitem{Haghighat:2013tka} 
  B.~Haghighat, C.~Kozcaz, G.~Lockhart and C.~Vafa,
  ``Orbifolds of M-strings,''
  Phys.\ Rev.\ D {\bf 89}, no. 4, 046003 (2014)
  [arXiv:1310.1185 [hep-th]].
  %%CITATION = ARXIV:1310.1185;%%
  %18 citations counted in INSPIRE as of 21 May 2015  
 
  %\cite{Kim:2013nva}
\bibitem{Kim:2013nva} 
  H.~C.~Kim, S.~Kim, S.~S.~Kim and K.~Lee,
  ``The general M5-brane superconformal index,''
  arXiv:1307.7660.
  %%CITATION = ARXIV:1307.7660;%%
  %22 citations counted in INSPIRE as of 07 Feb 2016
  
  %\cite{Kim:2014kta}
\bibitem{Kim:2014kta} 
  J.~Kim, S.~Kim, K.~Lee and J.~Park,
  ``Super-Yang-Mills theories on $S^{4} \times \mathbb{R}$,''
  JHEP {\bf 1408}, 167 (2014)
  %doi:10.1007/JHEP08(2014)167
  [arXiv:1405.2488 [hep-th]].
  %%CITATION = doi:10.1007/JHEP08(2014)167;%%
  %5 citations counted in INSPIRE as of 07 Feb 2016
  
  %\cite{Hohenegger:2015cba}
\bibitem{Hohenegger:2015cba} 
  S.~Hohenegger, A.~Iqbal and S.~J.~Rey,
  ``M-strings, monopole strings, and modular forms,''
  Phys.\ Rev.\ D {\bf 92}, no. 6, 066005 (2015)
  %doi:10.1103/PhysRevD.92.066005
  [arXiv:1503.06983 [hep-th]].
  %%CITATION = doi:10.1103/PhysRevD.92.066005;%%
  %6 citations counted in INSPIRE as of 09 Feb 2016
  
    %\cite{Hayashi:2015fsa}
\bibitem{Hayashi:2015fsa} 
  H.~Hayashi, S.~S.~Kim, K.~Lee, M.~Taki and F.~Yagi,
  ``A new 5d description of 6d D-type minimal conformal matter,''
  JHEP {\bf 1508}, 097 (2015)
  %doi:10.1007/JHEP08(2015)097
  [arXiv:1505.04439 [hep-th]].
  %%CITATION = doi:10.1007/JHEP08(2015)097;%%
  %9 citations counted in INSPIRE as of 09 févr. 2016


%\cite{Aharony:1997bh}
\bibitem{Aharony:1997bh}
  O.~Aharony, A.~Hanany and B.~Kol,
  ``Webs of (p,q) five-branes, five-dimensional field theories and grid diagrams,''
  JHEP {\bf 9801} (1998) 002
  [hep-th/9710116].
  %%CITATION = HEP-TH/9710116;%%
  %201 citations counted in INSPIRE as of 02 Jun 2015

%\cite{Gopakumar:1998ii}
\bibitem{Gopakumar:1998ii}
  R.~Gopakumar and C.~Vafa,
  ``M theory and topological strings. 1.,''
  hep-th/9809187.
  %%CITATION = HEP-TH/9809187;%%
  %298 citations counted in INSPIRE as of 30 May 2015

%\cite{Gopakumar:1998jq}
\bibitem{Gopakumar:1998jq}
  R.~Gopakumar and C.~Vafa,
  ``M theory and topological strings. 2.,''
  hep-th/9812127.
  %%CITATION = HEP-TH/9812127;%%
  %331 citations counted in INSPIRE as of 30 May 2015

%\cite{Leung:1997tw}
\bibitem{Leung:1997tw}
  N.~C.~Leung and C.~Vafa,
  ``Branes and toric geometry,''
  Adv.\ Theor.\ Math.\ Phys.\  {\bf 2} (1998) 91
  [hep-th/9711013].
  %%CITATION = HEP-TH/9711013;%%
  %175 citations counted in INSPIRE as of 02 juin 2015

%\cite{Hollowood:2003cv}
\bibitem{Hollowood:2003cv}
  T.~J.~Hollowood, A.~Iqbal and C.~Vafa,
  ``Matrix models, geometric engineering and elliptic genera,''
  JHEP {\bf 0803} (2008) 069
  [hep-th/0310272].
  %%CITATION = HEP-TH/0310272;%%
  %143 citations counted in INSPIRE as of 30 May 2015


%\cite{Aganagic:2002qg}
\bibitem{Aganagic:2002qg}
  M.~Aganagic, M.~Marino and C.~Vafa,
  ``All loop topological string amplitudes from Chern-Simons theory,''
  Commun.\ Math.\ Phys.\  {\bf 247} (2004) 467
  [hep-th/0206164].
  %%CITATION = HEP-TH/0206164;%%
  %117 citations counted in INSPIRE as of 12 juin 2015
  
%\cite{Iqbal:2002we}
\bibitem{Iqbal:2002we}
  A.~Iqbal,
  ``All genus topological string amplitudes and five-brane webs as Feynman diagrams,''
  hep-th/0207114.
  %%CITATION = HEP-TH/0207114;%%
  %57 citations counted in INSPIRE as of 12 juin 2015

%\cite{Iqbal:2003ix}
\bibitem{Iqbal:2003ix}
  A.~Iqbal and A.~K.~Kashani-Poor,
  ``Instanton counting and Chern-Simons theory,''
  Adv.\ Theor.\ Math.\ Phys.\  {\bf 7} (2004) 457
  [hep-th/0212279].
  %%CITATION = HEP-TH/0212279;%%
  %84 citations counted in INSPIRE as of 12 Jun 2015

%\cite{Iqbal:2003zz}
\bibitem{Iqbal:2003zz}
  A.~Iqbal and A.~K.~Kashani-Poor,
  ``SU(N) geometries and topological string amplitudes,''
  Adv.\ Theor.\ Math.\ Phys.\  {\bf 10} (2006) 1
  [hep-th/0306032].
  %%CITATION = HEP-TH/0306032;%%
  %90 citations counted in INSPIRE as of 12 juin 2015

%\cite{Eguchi:2003sj}
\bibitem{Eguchi:2003sj}
  T.~Eguchi and H.~Kanno,
  ``Topological strings and Nekrasov's formulas,''
  JHEP {\bf 0312} (2003) 006
  [hep-th/0310235].
  %%CITATION = HEP-TH/0310235;%%
  %71 citations counted in INSPIRE as of 12 Jun 2015

%\cite{Aganagic:2003db}
\bibitem{Aganagic:2003db}
  M.~Aganagic, A.~Klemm, M.~Marino and C.~Vafa,
  ``The Topological vertex,''
  Commun.\ Math.\ Phys.\  {\bf 254} (2005) 425
  [hep-th/0305132].
  %%CITATION = HEP-TH/0305132;%%
  %325 citations counted in INSPIRE as of 02 juin 2015
  

  %\cite{Awata:2005fa}
\bibitem{Awata:2005fa}
  H.~Awata and H.~Kanno,
  ``Instanton counting, Macdonald functions and the moduli space of D-branes,''
  JHEP {\bf 0505} (2005) 039
  [hep-th/0502061].
  %%CITATION = HEP-TH/0502061;%%
  %53 citations counted in INSPIRE as of 12 Jun 2015
  
  %\cite{Iqbal:2007ii}
\bibitem{Iqbal:2007ii}
  A.~Iqbal, C.~Kozcaz and C.~Vafa,
  ``The Refined topological vertex,''
  JHEP {\bf 0910} (2009) 069
  [hep-th/0701156].
  %%CITATION = HEP-TH/0701156;%%
  %171 citations counted in INSPIRE as of 30 May 2015

  
  %\cite{Taki:2007dh}
\bibitem{Taki:2007dh}
  M.~Taki,
  ``Refined Topological Vertex and Instanton Counting,''
  JHEP {\bf 0803} (2008) 048
  [arXiv:0710.1776 [hep-th]].
  %%CITATION = ARXIV:0710.1776;%%
  %40 citations counted in INSPIRE as of 12 juin 2015

%\cite{Awata:2008ed}
\bibitem{Awata:2008ed}
  H.~Awata and H.~Kanno,
  ``Refined BPS state counting from Nekrasov's formula and Macdonald functions,''
  Int.\ J.\ Mod.\ Phys.\ A {\bf 24} (2009) 2253
  [arXiv:0805.0191 [hep-th]].
  %%CITATION = ARXIV:0805.0191;%%
  %44 citations counted in INSPIRE as of 12 Jun 2015
  
%\cite{Iqbal:2012mt}
\bibitem{Iqbal:2012mt}
  A.~Iqbal and C.~Kozcaz,
  ``Refined Topological Strings and Toric Calabi-Yau Threefolds,''
  arXiv:1210.3016 [hep-th].
  %%CITATION = ARXIV:1210.3016;%%
  %12 citations counted in INSPIRE as of 12 Jun 2015
  
    %\cite{Iqbal:2004ne}
\bibitem{Iqbal:2004ne}
  A.~Iqbal and A.~K.~Kashani-Poor,
  ``The Vertex on a strip,''
  Adv.\ Theor.\ Math.\ Phys.\  {\bf 10} (2006) 317
  [hep-th/0410174].
  %%CITATION = HEP-TH/0410174;%%
  %61 citations counted in INSPIRE as of 10 juil. 2015

%\cite{Konishi:2006ev}
\bibitem{Konishi:2006ev}
  Y.~Konishi and S.~Minabe,
  ``Flop invariance of the topological vertex,''
  Int.\ J.\ Math.\  {\bf 19} (2008) 27
  [math/0601352 [math-ag]].
  %%CITATION = MATH/0601352;%%
  %14 citations counted in INSPIRE as of 10 juil. 2015

  
  \bibitem{Taki:2008hb} 
  M.~Taki,
  ``Flop Invariance of Refined Topological Vertex and Link Homologies,''
  arXiv:0805.0336 [hep-th].
  %%CITATION = ARXIV:0805.0336;%%
  %13 citations counted in INSPIRE as of 19 May 2015
  
   %\cite{Hayashi:2013qwa}
\bibitem{Hayashi:2013qwa}
  H.~Hayashi, H.~C.~Kim and T.~Nishinaka,
  ``Topological strings and 5d $T_N$ partition functions,''
  JHEP {\bf 1406} (2014) 014
  doi:10.1007/JHEP06(2014)014
  [arXiv:1310.3854 [hep-th]].
  %%CITATION = doi:10.1007/JHEP06(2014)014;%%
  %39 citations counted in INSPIRE as of 12 May 2016
  
  %\cite{Taki:2010bj}
\bibitem{Taki:2010bj} 
  M.~Taki,
  ``Surface Operator, Bubbling Calabi-Yau and AGT Relation,''
  JHEP {\bf 1107}, 047 (2011)
  doi:10.1007/JHEP07(2011)047
  [arXiv:1007.2524 [hep-th]].
  %%CITATION = doi:10.1007/JHEP07(2011)047;%%
  %48 citations counted in INSPIRE as of 07 Dec 2015
  

%\cite{Okuyama:2005gq}
\bibitem{Okuyama:2005gq} 
  K.~Okuyama,
  ``D1-D5 on ALE space,''
  JHEP {\bf 0512}, 042 (2005)
  doi:10.1088/1126-6708/2005/12/042
  [hep-th/0510195].
  %%CITATION = doi:10.1088/1126-6708/2005/12/042;%%
  %6 citations counted in INSPIRE as of 07 févr. 2016
  
  %\cite{Benini:2013nda}
\bibitem{Benini:2013nda}
  F.~Benini, R.~Eager, K.~Hori and Y.~Tachikawa,
  ``Elliptic genera of two-dimensional N=2 gauge theories with rank-one gauge groups,''
  Lett.\ Math.\ Phys.\  {\bf 104} (2014) 465
  [arXiv:1305.0533 [hep-th]].
  %%CITATION = ARXIV:1305.0533;%%
  %52 citations counted in INSPIRE as of 10 Jul 2015
  
  %\cite{Benini:2013xpa}
\bibitem{Benini:2013xpa}
  F.~Benini, R.~Eager, K.~Hori and Y.~Tachikawa,
  ``Elliptic Genera of 2d ${\mathcal{N}}$ = 2 Gauge Theories,''
  Commun.\ Math.\ Phys.\  {\bf 333} (2015) 3,  1241
  [arXiv:1308.4896 [hep-th]].
  %%CITATION = ARXIV:1308.4896;%%
  %54 citations counted in INSPIRE as of 10 juil. 2015
  
  %\cite{Gadde:2013dda}
\bibitem{Gadde:2013dda}
  A.~Gadde and S.~Gukov,
  ``2d Index and Surface operators,''
  JHEP {\bf 1403} (2014) 080
  [arXiv:1305.0266 [hep-th]].
  %%CITATION = ARXIV:1305.0266;%%
  %57 citations counted in INSPIRE as of 10 juil. 2015
  
  %\cite{Harvey:2014nha}
\bibitem{Harvey:2014nha}
  J.~A.~Harvey, S.~Lee and S.~Murthy,
  ``Elliptic genera of ALE and ALF manifolds from gauged linear sigma models,''
  JHEP {\bf 1502} (2015) 110
  [arXiv:1406.6342 [hep-th]].
  %%CITATION = ARXIV:1406.6342;%%
  %7 citations counted in INSPIRE as of 10 juil. 2015
  
  %\cite{Honda:2015yha}
\bibitem{Honda:2015yha}
  M.~Honda and Y.~Yoshida,
  ``Supersymmetric index on $T^2$ x $S^2$ and elliptic genus,''
  arXiv:1504.04355 [hep-th].
  %%CITATION = ARXIV:1504.04355;%%
  %3 citations counted in INSPIRE as of 10 juil. 2015
  


%\cite{Sulkowski:2006jp}
%\bibitem{Sulkowski:2006jp}
  %P.~Sulkowski,
  %``Crystal model for the closed topological vertex geometry,''
  %JHEP {\bf 0612} (2006) 030
  %[hep-th/0606055].
  %%CITATION = HEP-TH/0606055;%%
  %21 citations counted in INSPIRE as of 02 juin 2015
  


%\cite{Dijkgraaf:2006um}
%\bibitem{Dijkgraaf:2006um}
  %R.~Dijkgraaf, C.~Vafa and E.~Verlinde,
  %``M-theory and a topological string duality,''
  %hep-th/0602087.
  %%CITATION = HEP-TH/0602087;%%
  %61 citations counted in INSPIRE as of 01 Jun 2015

\end{thebibliography}
\end{document}